# Probing Carrier Transport and Structure-property Relationship of Highly Ordered Organic Semiconductors at Two-dimensional Limit

Yuhan Zhang,[1,†] Jingsi Qiao,[2,†] Si Gao,[3] Fengrui Hu,[4] Daowei He,[1] Bing Wu,[1] Ziyi Yang,[1] Bingchen Xu,[1] Yun Li,[1] Yi Shi,[1,*] Wei Ji,[2,5*] Peng Wang,[3] Xiaoyong Wang,[4] Min Xiao,[4,6] Hangxun Xu,[7] Jian-Bin Xu,[8,1*] and Xinran Wang[1,*]

[1]*National Laboratory of Solid State Microstructures, School of Electronic Science and Engineering, and Collaborative Innovation Center of Advanced Microstructures, Nanjing University, Nanjing 210093, China*

[2]*Department of Physics and Beijing Key Laboratory of Optoelectronic Functional Materials & Micro-nano Devices, Renmin University of China, Beijing 100872, China.*

[3]*College of Engineering and Applied Sciences, Nanjing University, Nanjing 210093, China*

[4]*School of Physics, Nanjing University, Nanjing 210093, China*

[5]*Department of Physics and Astronomy, Shanghai Jiao Tong University, Shanghai 200240, China and Collaborative Innovation Center of Advanced Microstructures, Nanjing 210093, China*

[6]*Department of Physics, University of Arkansas, Fayetteville, AR72701, USA*

[7]*CAS Key Laboratory of Soft Matter Chemistry, Department of Polymer Science and Engineering, University of Science and Technology of China, Hefei 230026, China*

[8]*Department of Electronic Engineering and Materials Science and Technology Research Center, The Chinese University of Hong Kong, Hong Kong SAR, China*

* Correspondence to X. W. (xrwang@nju.edu.cn), J.-B. X. (jbxu@ee.cuhk.edu.hk), Y. S. (yshi@nju.edu.cn) and W. J. (wji@ruc.edu.cn).

† *These authors contribute equally to this work.*

# Abstract

One of the basic assumptions in organic field-effect transistors, the most fundamental device unit in organic electronics, is that charge transport occurs two-dimensionally in the first few molecular layers near the dielectric interface. Although the mobility of bulk organic semiconductors has increased dramatically, direct probing of intrinsic charge transport in the two-dimensional limit has not been possible due to excessive disorders and traps in ultrathin organic thin films. Here, highly ordered mono- to tetra-layer pentacene crystals are realized by van der Waals (vdW) epitaxy on hexagonal BN. We find that the charge transport is dominated by hopping in the first conductive layer, but transforms to band-like in subsequent layers. Such abrupt phase transition is attributed to strong modulation of the molecular packing by interfacial vdW interactions, as corroborated by quantitative structural characterization and density functional theory calculations. The structural modulation becomes negligible beyond the second conductive layer, leading to a mobility saturation thickness of only ~3nm. Highly ordered organic ultrathin films provide a platform for new physics and device structures (such as heterostructures and quantum wells) that are not possible in conventional bulk crystals.

Organic field-effect transistors (OFETs) offer unique advantages of low cost, lightweight and flexibility and are widely used in electronics and display industry. While the mobility of bulk organic semiconductors has increased dramatically [1-3], an outstanding issue is to directly examine the structure-property relationship at the semiconductor-dielectric interface [4], where charge transport actually occurs [5-7]. Ultrathin organic semiconductors down to few-nanometre thickness are often dominated by traps and disorders and far away from intrinsic transport regime [8-10]. Another challenge in organic electronics is the development of layer-by-layer epitaxy with the precision similar to molecular beam epitaxy in their inorganic counterparts [11]. These challenges may be alleviated if molecular crystals are processed into large-area, highly crystalline monolayers. Such 2D form factor will also bring about new applications such as nanoporous membranes and insulating dielectrics [12, 13]. Several recent breakthroughs in various types of 2D organic materials such as polymers [14, 15], oligomers [16] and covalent organic frameworks [17] have already shown great promises along this direction. However, one of the most fundamental questions regarding the nature of charge transport at the 2D limit has not been addressed. In this work, we study the benchmark molecule pentacene epitaxially crystallized on BN substrate because of its high mobility and simple structure to model. The highly clean system allows us to provide the first definitive scenario of how molecular packing and charge transport are modulated near the interface, without being dominated by extrinsic factors. Our results suggest the possibility of band-like transport in organic materials even at the monolayer limit. This hybrid structure can also serve as a generic platform to study the intrinsic electrical and optical properties of organic semiconductors down to monolayer.

Few-layer pentacene crystals were epitaxially grown on mechanically exfoliated hexagonal BN by vapour transport method in a tube furnace [see 18]. The reason to choose BN as the epitaxial substrate is two-fold. First, it is atomically flat with no dangling bonds and low density of impurities, crucial to realize high-quality single-crystal pentacene films. Second, BN has low dielectric constant, which, according to the Fröhlich polaron picture [19-20], should give a weak polaronic coupling. The growth proceeded in a layer-by-layer fashion with clear anisotropy. The frequent appearance of well-defined crystal facets (Fig. 1f, Supplementary Fig. S2b, S2c) [18] indicated that the pentacene was crystalline, as later confirmed by atomic force microscopy (AFM) and transmission electron microscopy (TEM). As schematically illustrated in Fig. 1a, the molecular packing is very different near the dielectric interface. The average thickness of the wetting layer (WL, also referred to as interfacial layer in the literature), the first conducting layer (1L) and the second conducting layer (2L) is 0.5nm, 1.14nm, and 1.58nm, respectively (Figs. 1b, d-f). The subsequent layers have the same height and molecular packing as 2L. The small thickness of WL suggests that the molecules adopt the face-on configuration (Supplementary Fig. S15 [18]), similar to that of pentacene on graphite [21] and on metal [22]. The thickness of 2L is consistent with the thin-film phase of pentacene [23]. However, 1L is clearly a new polymorph, whose reduced height compared to 2L suggests more tilted molecular packing.

Further structural information of the pentacene layers was gained by high-resolution AFM. We found that both 1L and 2L were highly crystalline with the typical herringbone-like packing in the (001) plane (Fig. 2a, Supplementary Fig. S3 [18]). The difference of lattice constants between 1L and 2L was not obvious upon initial inspection of the AFM images, but was unambiguously revealed by statistical analysis from a number of samples. As shown in Fig. 2c, the lattice constants along $a$- and $b$-axis were 6.23±0.07Å and 7.77±0.08Å (6.03±0.05Å and 7.76±0.05Å) for 1L (2L). Due to the reduced height of 1L, the unit cell expanded significantly by 0.2Å (or 3.3%) along $a$-axis, but little expansion was observed along $b$-axis. More pronounced difference had also been observed along $a$-axis when comparing the bulk and thin-film phase of pentacene [23].

The thermal drift of AFM under ambient conditions may introduce sub-nanometre-scale uncertainties that cause the finite width of distributions in Fig. 2c. Therefore, we also performed TEM characterization to crosscheck with AFM. Supplementary Fig. S4a [18] shows a typical low-magnification TEM image of few-layer pentacene on BN. Selected-area electron diffraction (SAED) was typically performed over an ~10μm$^2$ area and exhibited a single set of diffraction patterns from pentacene (Fig. 2b). Using the diffraction spots from BN as references, we determined that the lattice constants of the 2L pentacene [24] were 5.98Å±0.09Å and 7.61Å±0.13Å along $a$- and $b$-axes, and the angle between them is 88.25°±1.22°. Together with the height measurements, we conclude that the structure of 2L is consistent with the thin-film phase of pentacene. Statistical analysis of the SAED patterns showed a sharp peak near 16° between pentacene (010) and BN (100) (Supplementary Fig. S4c [18]), indicating that pentacene had a quasi-epitaxial relationship with BN [12]. The quasi-epitaxy nature was due to weak vdW interactions and incommensurability between pentacene and BN.

Our structure measurements were further supported by *ab initio* DFT calculations [18]. Supplementary Fig. S17[18]shows the optimized molecular packing for 1L and 2L, whose lattice constants are in good agreement with experimental values within 1.5%. The packing of pentacene in each layer depends critically on the competition between intralayer and interlayer interactions. The former favours upright packing, while the latter favours face-on packing [25]. In WL, the pentacene molecules adopt a face-on configuration with the long axis along the $[1\bar{1}2]$ direction of BN, because of their strong interactions of 2.35 eV/molecule (Supplementary Fig. S15, Table S1 [18]). In 2L, on the other hand, the substantially reduced interlayer interaction (less than 0.3 eV/molecule) leads to the thin-film-like packing. 1L is obviously a transition between the two extremes. The leaning of molecules in 1L occurs primarily along the *b*-axis to maximize the π-π stacking between WL and 1L. The molecules then reorient their shorter axis more parallel to the *a*-axis, which leads to closer distance and thus more repulsion between neighboring molecules along the same direction (Supplementary Fig. S17 [18]). To release this repulsion, the unit cell of 1L mainly expands along *a*-axis, as observed experimentally.

The pristine nature of bulk organic crystals is often manifested by the anisotropy in their optical and electrical properties [26-28]. We carried out polarization-dependent absorption and photoluminescence (PL) on 1L and 2L samples to demonstrate such anisotropy. Both absorption and PL exhibited clear and uniform modulations with a period of ~180° (Figs. 2e, 2f, Supplementary Figs. S5, S6 [18]), presumably tailored by the crystal symmetry. The direction of the highest (lowest) PL intensity was assigned to the *a*-axis (*b*-axis) of the pentacene crystals [29]. Single-crystalline 1L with lateral size up to ~60μm has been observed (Supplementary Fig. S5 [18]), only limited by the size of BN. We note that the anisotropy of PL had only been indirectly observed in the highest quality pentacene single crystals by ellipsometry and electron energy-loss spectroscopy [29]. The PL of 1L and 2L samples composed of two prominent peaks, centred at 2.16eV and 2.29eV for 1L and 2.13eV and 2.25eV for 2L, as well as a small peak near 2.4eV (Supplementary Fig. S6 [18]). The splitting of ~0.12eV between the two main peaks can be attributed to Davydov splitting from the two non-equivalent molecules in a unit cell. Compared to the free exciton state in pentacene thin-films and monolayers [30, 31], the most striking feature was the large blue shift of exciton energy (or equivalently, the reduction of exciton binding energy) on the order of 0.3eV. A rough estimate gives the exciton radius of several nanometres [32], indicating their highly delocalized (or Wannier-Mott) nature likely due to the good crystallinity of the pentacene.

Next we focus on the thickness-dependent electrical transport in pentacene crystals using backgated OFET geometry. We found that WL was not conducting with the detection limit of our instruments, consistent with the absence of intralayer π-π stacking. In the following, we focus our discussions on one representative device from 1L, 2L and 3L, but the data were qualitatively and consistently reproduced in other devices [18]. We also checked the reversibility of our devices after thermal cycling to ensure that the observations were fully repeatable and not due to artefacts (Supplementary Fig. S14 [18]).

Fig. 3a and Supplementary Fig. S7b [18] show the room-temperature transfer ($I_{ds}$-$V_g$) and output ($I_{ds}$-$V_{ds}$) characteristics of a 1L device with on/off ratio ~$10^8$. Several textbook features of high-quality OFET were observed in spite of the monolayer channel thickness [8]: exceptional linearity of transfer characteristics in the linear (low bias) regime, high field-effect mobility μ=1.6cm$^2$/Vs (all the field-effect mobilities in this paper are measured from linear regime unless otherwise noted), Ohmic contact (Supplementary Fig. S7c [18]), nearly zero threshold voltage ($V_{th}$=1.5V, corresponding to a density of deep traps ~$10^{11}$ cm$^{-2}$), small subthreshold swing (*SS*=450mV/decade) and little hysteresis (Supplementary Fig. S13a [18]).

Further insights of charge transport were inferred by temperature-dependent electrical measurements. We found that all the 1L devices exhibited insulating behavior, along with increasing nonlinearity of the $I_{ds}$-$V_g$ characteristics at low temperature (Fig. 3b). The transfer characteristics in the linear regime could be well described by a power-law relationship $(V_g - V_{th})^\gamma$ where the exponent adopted an inverse scaling with temperature $\gamma = \frac{T_0}{T}$. These features are signatures of 2D hopping transport [33, 34]. From the linear fitting of the power exponent $\gamma$, we can extract the Urbach energy of the localized states $T_0$=331K (Fig. 3b inset). The Urbach energy is much smaller than disordered 2D organic semiconductors [34], and comparable to the best value for conjugated polymers [35]. By adopting the fitting procedure in Ref. [34] [18], we further deduced the localization length $\alpha^{-1}\approx$0.82nm. Another 1L device showed similar $\alpha^{-1}\approx$0.94nm (Supplementary Fig. S9 [18]). As we shall show later with DFT calculations, the localization length ~1nm is a natural result of the molecular packing in 1L.

A more surprising observation comes from 2L devices, which exhibit band-like transport [27]. The room-temperature field-effect mobility was typically ~3 cm$^2$/Vs (Fig. 3e), slightly higher than the 1L devices. However, the difference in mobility became very dramatic (up to 50 times) at low temperature as the mobility of 2L devices improved as $T$ was lowered, consistent with band transport and lack of localization (Figs. 3d-f, Supplementary Fig. S10c [18]). The low-temperature field-effect mobility reached up to 5.2cm$^2$/Vs, far exceeding pentacene polycrystalline thin-film devices at similar temperatures [10, 36]. The band-like behavior could extend down to our base temperature of 50K at high carrier density (Fig. 3f). At low carrier density, the weakly insulating regime at low temperatures pointed out still finite density of shallow traps. But the large temperature and carrier density window for band-like transport suggested rather small Urbach energy of the trap states. Indeed, Arrhenius-type fitting of the mobility in the low temperature regime gave an estimate of the Urbach energy on the order of a few meV, comparable to the thermal energy at 50K.

The strong modulation of charge transport near the interface is unlikely from extrinsic factors such as impurities because we adopt the same material, substrate and growth procedure for 1L and 2L. To understand the molecular origin behind it, we carried out DFT calculations [18]. Fig. 4 visualizes the molecular orbitals of the inter-molecular bonding states for 1L (1L-B) and 2L (2L-B), which are responsible for the hole conduction in both layers (Supplementary Figs. S19, S20 [18]). In 2L, the orbital overlaps horizontally, resulting in a fully extended density of states along *a*- and *b*-axes that is likely responsible for the band-like transport. The more tilted molecular packing in 1L, however, substantially modifies the spatial distribution of the bonding states 1L-B, so much so that the orbitals only span for five molecules along *b*-axis. Therefore, a hole in 1L can only travel for ~1nm before localized near the WL-1L interface (Supplementary Fig. S20 [18]). The localization length is in excellent agreement with experimental value without any adjustable parameters. The nature of the molecular orbitals can be more clearly visualized within *a-b* plane (Figs. 4b, c), where 2L-B clearly forms a continuous 2D network, while 1L-B appears disconnected and localized in both directions. We believe this is mainly responsible for the distinct transport behavior in 1L and 2L. Additional reasons may include the different interlayer coupling. The transfer integral between 1L and WL has a similar magnitude to that between the adjacent 1L molecules (Supplementary Table S2[18]). This strong interlayer coupling, acting as a disorder perturbation to 1L, may cause further localization of charge carriers. The transport in 2L is nearly unperturbed from 1L, ascribed to the much smaller transfer integrals between them.

Mobility saturation is also an important issue in OFETs. Many early studies suggested that the saturation thickness was material-dependent, and was ~6 layers (or ~10nm) for polycrystalline pentacene [7]. However, the molecular understanding of the saturation thickness has remained unclear. We note that the polycrystalline thin films in those early works have high density of defects and domain boundaries, which may facilitate interlayer vertical transport. To explore this issue, we also studied 3L devices (Supplementary Fig. S12 [18]) We observed the same qualitative transport behavior as 2L devices, including room-temperature field-effect mobility ~2-3cm$^2$/Vs and band-like transport. We thus conclude that it only takes two conducting layers (or ~3nm) to reach mobility saturation in our epitaxial pentacene. This is because the modulation of molecular packing and charge transport by the substrate is already negligible beyond 2L. The small saturation thickness should be a generic attribute of high-quality layered organic semiconductors.

In conclusion, we demonstrate that vdW epitaxy of high-quality, few-layer molecular crystals can provide a powerful platform to explore their intrinsic charge transport down to monolayer. Our results clearly show that interfacial modulation by vdW forces is an effective means to engineer the properties of organic semiconductors. We believe the highly ordered 2D molecular crystals with clean transport and excitonic properties as demonstrated here may lead to new quantum phenomena that have thus far been prevented by disorders, and to new device structures based on precise assembly of organic layers [37]. To this end, Alves et al. [38] and Gutierrez et al. [39] showed that charge-transfer states formed at the interface of two bulk organic crystals could be exploited for device applications. We believe the concept developed here will be able to realize more versatile structures such as planar, vertical heterostructures and quantum wells.

**Acknowledgements**

This work was supported in part by National Key Basic Research Program of China 2013CBA01604, 2015CB921600, 2012CB932704, 2015CB351900; National Natural Science Foundation of China 61325020, 61261160499, 11274154, 61521001, 11274380, 91433103, 61229401; Research Grant Council of Hong Kong SAR N_CUHK405/12; MICM Laboratory Foundation 9140C140105140C14070, a project funded by the Priority Academic Program Development of Jiangsu Higher Education Institutions, "Jiangsu Shuangchuang" program and 'Jiangsu Shuangchuang Team'

**References:**

[1]  H. Li, G. Giri, J. B. H. Tok, and Z. Bao, MRS Bulletin **38**, 34 (2013).
[2]  C. Wang, H. Dong, W. Hu, Y. Liu, and D. Zhu, Chem. Rev. **112**, 2208 (2012).
[3]  G. Giri *et al.*, Nature **480**, 504 (2011).
[4]  N. A. Minder, S. F. Lu, S. Fratini, S. Ciuchi, A. Facchetti, and A. F. Morpurgo, Adv. Mater. **26**, 1254 (2014).
[5]  A. Dodabalapur, L. Torsi, and H. E. Katz, Science **268**, 270 (1995).
[6]  F. Dinelli, M. Murgia, P. Levy, M. Cavallini, F. Biscarini, and D. M. de Leeuw, Phys. Rev. Lett. **92**, 116802 (2004).
[7]  R. Ruiz, A. Papadimitratos, A. C. Mayer, and G. G. Malliaras, Adv. Mater. **17**, 1795 (2005).
[8]  V. Podzorov, MRS Bulletin **38**, 15 (2013).
[9]  H. Sirringhaus, Adv. Mater. **17**, 2411 (2005).
[10] J. W. Wang and C. Jiang, Org. Electr. **16**, 164 (2015).
[11] S. R. Forrest, Chem. Rev. **97**, 1793 (1997).
[12] X. D. Zhuang, Y. Y. Mai, D. Q. Wu, F. Zhang, and X. L. Feng, Adv. Mater. **27**, 403 (2015).
[13] H. Moon *et al.*, Nature Mater. **14**, 628 (2015).
[14] P. Kissel, D. J. Murray, W. J. Wulftange, V. J. Catalano, and B. T. King, Nat. Chem. **6**, 774 (2014).
[15] M. J. Kory, M. Worle, T. Weber, P. Payamyar, S. W. van de Poll, J. Dshemuchadse, N. Trapp, and A. D. Schluter, Nat. Chem. **6**, 779 (2014).
[16] D. He *et al.*, Nat. Commun. **5**, 5162 (2014).
[17] J. W. Colson, A. R. Woll, A. Mukherjee, M. P. Levendorf, E. L. Spitler, V. B. Shields, M. G. Spencer, J. Park, and W. R. Dichtel, Science **332**, 228 (2011).
[18] See Supplemental Material [url], which includes Refs. [40-53].
[19] I. N. Hulea, S. Fratini, H. Xie, C. L. Mulder, N. N. Iossad, G. Rastelli, S. Ciuchi, and A. F. Morpurgo, Nat. Mater. **5**, 982 (2006).
[20] J. Veres, S. D. Ogier, S. W. Leeming, D. C. Cupertino, and S. M. Khaffaf, Adv. Funct. Mater. **13**, 199 (2003).
[21] Y. Harada, H. Ozaki, and K. Ohno, Phys. Rev. Lett. **52**, 2269 (1984).
[22] Y. L. Wang, W. Ji, D. X. Shi, S. X. Du, C. Seidel, Y. G. Ma, H. J. Gao, L. F. Chi, and H. Fuchs, Phys. Rev. B **69**, 075408 (2004).
[23] S. C. B. Mannsfeld, A. Virkar, C. Reese, M. F. Toney, and Z. Bao, Adv. Mater. **21**, 2294 (2009).

[24] We believe the diffractions spots mainly come from 2L and above, because the electron reflections from IL and 1L are considerably weaker than those from the rest of the pentacene multi-layers and hence not able to be recorded at the current imaging condition with a short acquisition time and low incident electron dose.
[25] K. Kim, E. J. G. Santos, T. H. Lee, Y. Nishi, and Z. Bao, Small **11**, 2037 (2015).
[26] V. C. Sundar, J. Zaumseil, V. Podzorov, E. Menard, R. L. Willett, T. Someya, and J. A. Rogers, Science **303**, 1644 (2004).
[27] V. Podzorov, E. Menard, A. Borissov, V. Kiryukhin, J. A. Rogers, and M. E. Gershenson, Phys. Rev. Lett. **93**, 086602 (2004).
[28] D. Y. Zang, F. F. So, and S. R. Forrest, Appl. Phys. Lett. **59**, 823 (1991).
[29] M. Grobosch, R. Schuster, T. Pichler, M. Knupfer, and H. Berger, Phys. Rev. B **74**, 155202 (2006).
[30] R. He, X. Chi, A. Pinczuk, D. V. Lang, and A. P. Ramirez, Appl. Phys. Lett. **87**, 211117 (2005).
[31] R. He, N. G. Tassi, G. B. Blanchet, and A. Pinczuk, Appl. Phys. Lett. **96**, 263303 (2010).
[32] Considering the gap between HOMO and LUMO in pentacene is 2.2-2.3eV (Phys. Rev. Lett. 93, 086802 (2004), Phys. Rev. B 71, 081202R (2005)), the observed exciton binding energy is on the order of 0.1eV. Using a dielectric constant of 2.5 for pentacene (Phys. Rev. Lett. 98, 037402 (2007)), we can estimate the exciton radius of 5.7nm.
[33] M. C. J. M. Vissenberg and M. Matters, Phys. Rev. B **57**, 12964 (1998).
[34] J. J. Brondijk, W. S. C. Roelofs, S. G. J. Mathijssen, A. Shehu, T. Cramer, F. Biscarini, P. W. M. Blom, and D. M. de Leeuw, Phys. Rev. Lett. **109**, 056601 (2012).
[35] D. Venkateshvaran *et al.*, Nature **515**, 384 (2014).
[36] A. R. Brown, C. P. Jarrett, D. M. deLeeuw, and M. Matters, Synth. Met. **88**, 37 (1997).
[37] A. K. Geim and I. V. Grigorieva, Nature **499**, 419 (2013).
[38] H. Alves, A. S. Molinari, H. X. Xie, and A. F. Morpurgo, Nat. Mater. **7**, 574 (2008).
[39] I. G. Lezama, M. Nakano, N. A. Minder, Z. H. Chen, F. V. Di Girolamo, A. Facchetti, and A. F. Morpurgo, Nat. Mater. **11**, 788 (2012).
[40] W. Regan, N. Alem, B. N. Alemán, B. Geng, C. a. l. Girit, L. Maserati, F. Wang, M. Crommie, and A. Zettl, *Appl. Phys. Lett.* **96**, 113102 (2010).
[41] G. E. Pike and C. H. Seager, *Phys. Rev. B* **10**, 1421 (1974).
[42] P. E. Blochl, *Phys. Rev. B* **50**, 17953 (1994).
[43] G. Kresse and D. Joubert, *Phys. Rev. B* **59**, 1758 (1999).
[44] G. Kresse and J. Furthmuller, *Phys. Rev. B* **54**, 11169 (1996).
[45] M. Dion, H. Rydberg, E. Schroder, D. C. Langreth, and B. I. Lundqvist, *Phys. Rev. Lett.* **92**, 246401(2004).
[46] K. Lee, E. D. Murray, L. Z. Kong, B. I. Lundqvist, and D. C. Langreth, *Phys. Rev. B* **82**, 081101(2010).

[47] J. Klimes, D. R. Bowler, and A. Michaelides, *J. Phys. Cond. Mat.* **22**, 074203 (2010).
[48] J. Klimes, D. R. Bowler, and A. Michaelides, *Phys. Rev. B* **83**,195131 (2011).
[49] J. Heyd, G. E. Scuseria, and M. Ernzerhof, *J. Chem. Phys.* **118**, 8207 (2003).
[50] J. Heyd, G. E. Scuseria, and M. Ernzerhof, *J. Chem. Phys.* **124**, 219906 (2006).
[51] W. Q. Deng and W. A. Goddard, *J. Phys. Chem. B* **108**, 8614 (2004).
[52] S. Schiefer, M. Huth, A. Dobrinevski, and B. Nickel, *J. Am. Chem. Soc.* **129**, 10316 (2007).
[53] P. B. Paramonov, V. Coropceanu, and J. L. Bredas, *Phys. Rev. B* **78**,041403(2008).

**FIGURE CAPTIONS**

FIG.1. Epitaxial growth of 2D pentacene crystals on BN. (a) Schematic illustration of the molecular packing of WL, 1L and 2L within *b-c* plane. (b) Histogram distribution of the thickness of WL, 1L and 2L, each taken from over 10 samples. (c) Raman spectrum of the pentacene crystals on BN, taken from a 2L sample. (d-f) AFM images of WL, 1L and 2L pentacene crystals on BN, respectively. The layer numbers are marked on each image. Insets show the height profiles along the dashed lines. The scale bars are 2μm.

FIG.2. Characterizations of 2D pentacene crystals. (a) High-resolution AFM images of a 1L sample. The unit cell is marked. The scale bar is 1nm. (b) SAED pattern from a few-layer pentacene sample (Supplementary Fig. S4a [18] shows the low-magnification TEM image of the sample). Blue circles mark the BN (100) directions and green circles mark the pentacene (110), (120) and (020) directions. (c) Histogram of lattice constants of 1L (upper panel) and 2L (lower panel) pentacene

crystals. Blue and red lines represent *a*- and *b*-axes, respectively. Black lines show the best Gaussian fittings. (d) Molecular packing of 1L (upper panel) and 2L (lower panel) pentacene within *a-b* plane. The unit cell is marked by the dashed rectangular box. (e) Polarization-dependent PL spectrum from a 2L sample. The black (blue) curve is taken when the PL intensity is strongest (weakest). Red lines are the fitting results with three Gaussian peaks (Supplementary Fig. S6d [18]). Inset is the spatially resolved PL image of the 2L sample, showing excellent uniformity. The scale bar is 3μm. (f) Normalized PL intensity of the 2L sample in **e** as a function of linear polarization angle.

FIG.3. Temperature-dependent electrical transport of 1L and 2L pentacene OFETs. (a) Room temperature $I_{ds}$-$V_g$ characteristics ($V_{ds}$=-2V, black line) and the extracted field-effect mobility as a function of $V_g$ (red symbols) of a 1L device. Inset shows the optical microscope image of the device. The scale bar is 20μm. (b) $I_{ds}$-$V_g$ characteristics at different temperatures of the same device under $V_{ds}$=-2V (symbols), plotted on a double logarithmic scale. From top to bottom, *T*=300K, 250K and 140K respectively. The lines are power-law fitting results. Inset shows the extracted power exponent as a function of 1000/*T* (symbols). The linear fitting crosses the origin (line), consistent with the 2D hopping mechanism. $T_0$=331K is derived from the linear fitting. (c) Experimental (symbols) and calculated (lines) mobility as a function of 1000/*T* under $V_g$=-30V (purple), -20V (blue) and -10V (orange). The calculations are done with the following parameters: $T_0=331K$, $\sigma_0=1.3\times10^6$S/m, $\alpha^{-1}=8.2$Å. (d) $I_{ds}$-$V_g$ characteristics ($V_{ds}$=-2V) at different temperatures of a 2L device. (e) The extracted mobility as a function of $V_g$ at the same temperatures as in (d). (f) Mobility as a function of temperature under $V_g$=-20V (orange), -35V (blue) and -50V (purple).

FIG.4. Visualized molecular orbitals of the inter-molecular bonding states for 1L and 2L. (a) Side views of the molecular orbitals for 1L-B (blue) and 2L-B (red) in the *b-c* plane, illustrated by isosurface contours of 0.00015 *e*/Bohr³. (b-c) Top views of the molecular orbitals in the slices cleaved in the *a-b* plane for 1L-B (b) and 2L-B (c). Yellow dashed lines in (a) indicate the positions of the slices. The top phenyl group of the pentacene molecules are drawn in (b) and (c).

Figure 1:

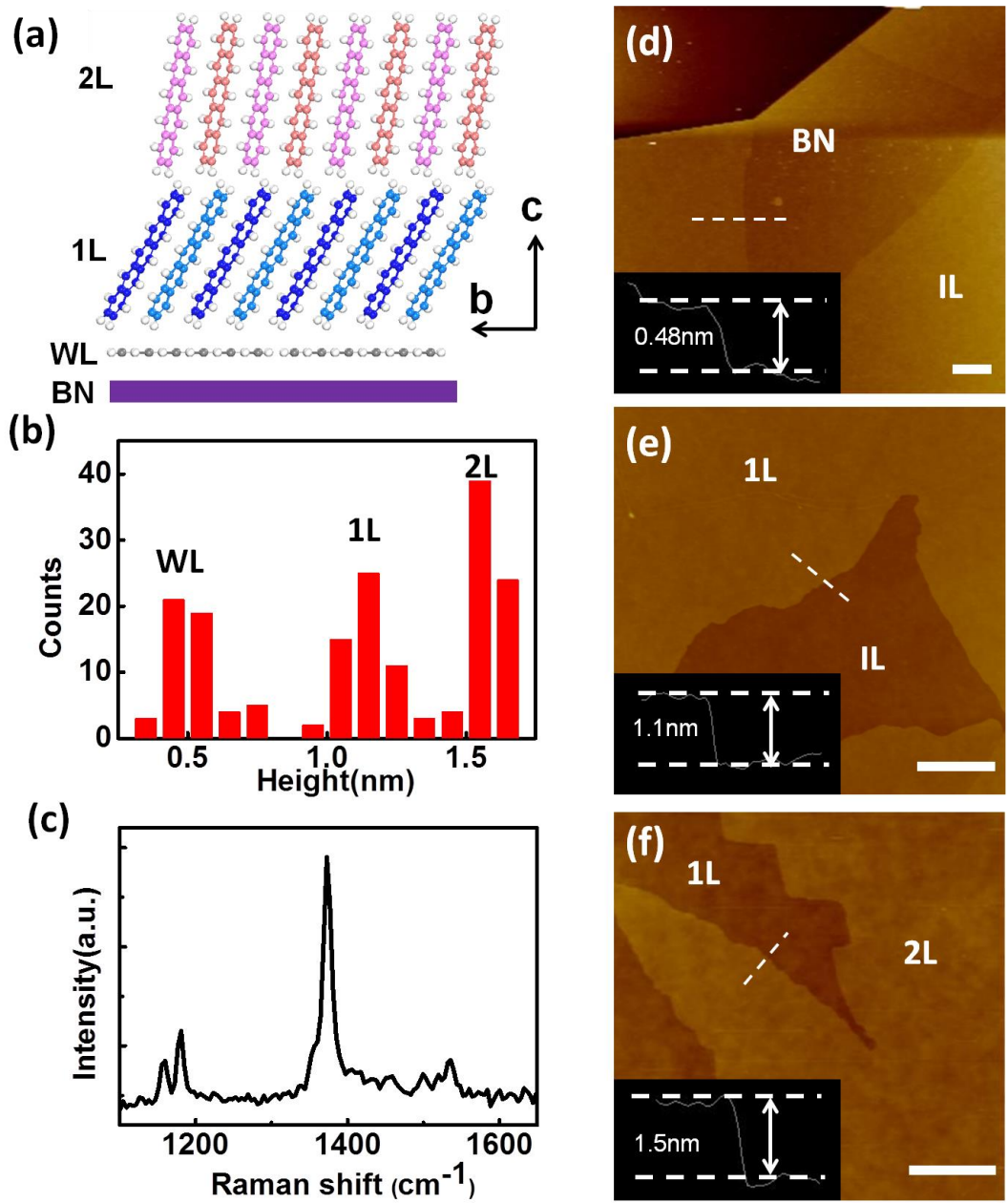

Figure 2:

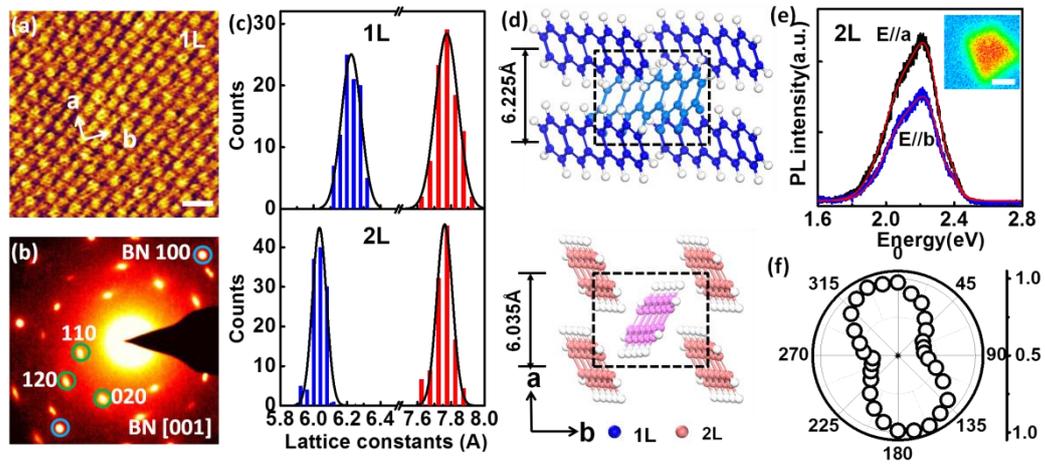

Figure 3

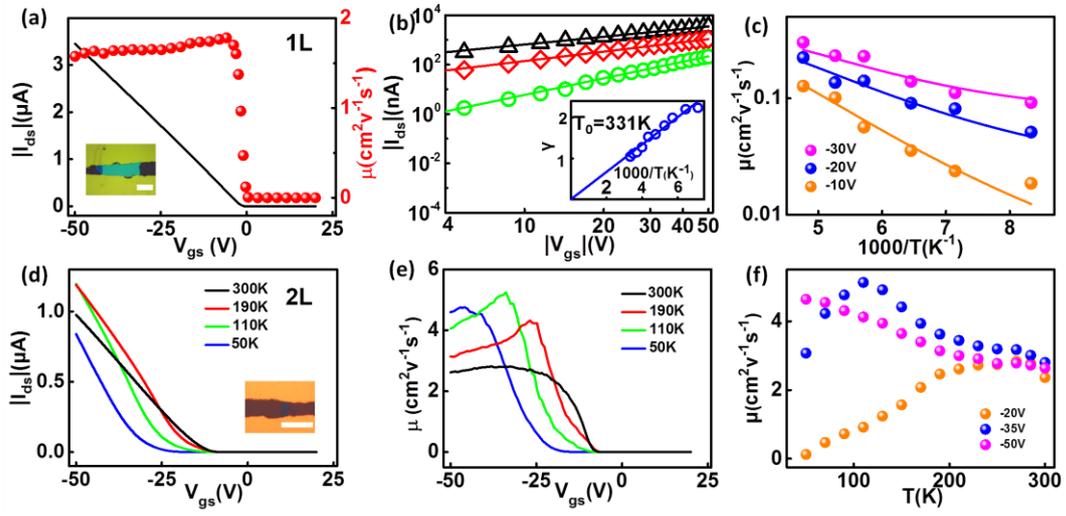

Figure 4:

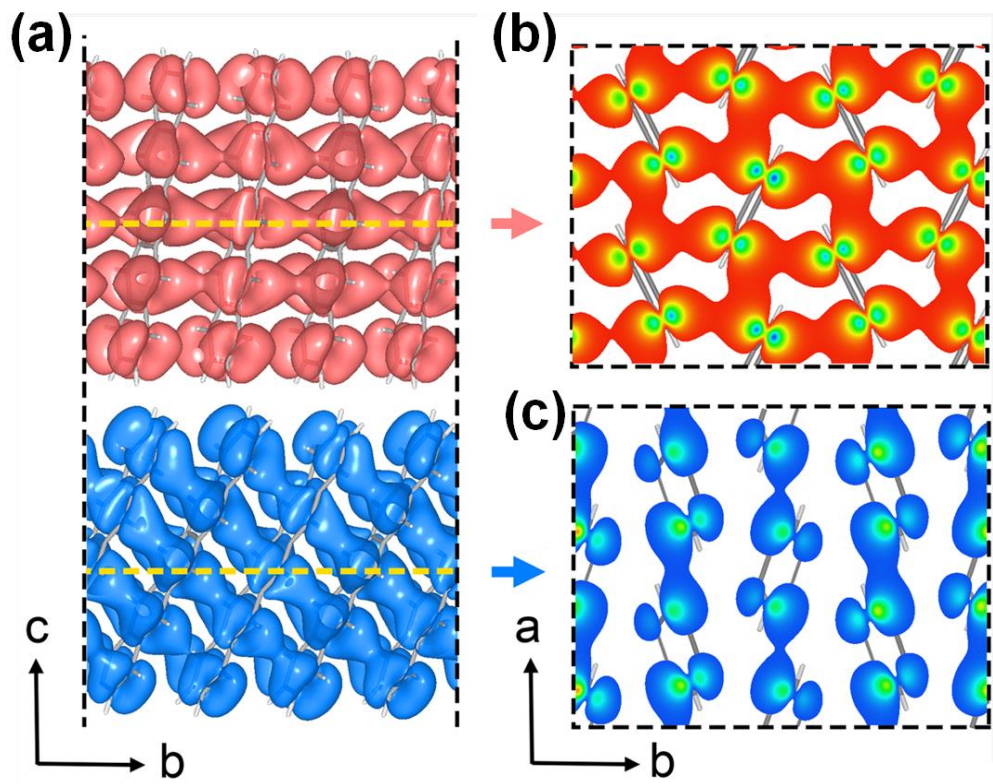

**Probing Carrier Transport and Structure-property Relationship of Highly Ordered Organic Semiconductors at Two-dimensional Limit**

Yuhan Zhang,[1,†] Jingsi Qiao,[2,†] Si Gao,[3] Fengrui Hu,[4] Daowei He,[1] Bing Wu,[1] Ziyi Yang,[1] Bingchen Xu,[1] Yun Li,[1] Yi Shi,[1,*] Wei Ji,[2,5*] Peng Wang,[3] Xiaoyong Wang,[4] Min Xiao,[4,6] Hangxun Xu,[7] Jian-Bin Xu,[8,1*] and Xinran Wang[1,*]

**Supplementary Information**

**1. Methods**

**Epitaxial growth of few-layer pentacene crystals on BN**

We used mechanically exfoliated few-layer BN sheet on 285nm-thick $SiO_2$ on Si ($SiO_2$/Si) as the growth substrate without further annealing. Before growth, the BN sheet was characterized by optical microscope and AFM to obtain the topological information. The epitaxial growth of pentacene was carried out in a home-built tube furnace. We put the pentacene powder (purchased from Sigma Aldrich without further purification) at the center of the furnace and the BN sheets a few inches downstream and used a turbo molecular pump to evacuate the quartz tube to $\sim 4 \times 10^{-6}$ Torr. We then heated up the furnace to 130~160 °C to grow pentacene crystals. The number of pentacene layers was controlled by the source temperature, the substrate position and growth time. When the growth was finished, we turned off the furnace and let the sample cool down to room temperature under vacuum.

**Details of AFM, TEM, Raman, polarization-dependent absorption and PL measurements**

AFM (both regular and high-resolution) was performed by an Asylum Cypher under ambient conditions. We used Asylum ARROW UHF tips for high-resolution AFM.

TEM and SAED images were collected using a FEI Tecnai F20 working under 200kV. To prepare for the TEM samples, few-layer BN sheets exfoliated on $SiO_2$/Si substrate were transferred onto Ni TEM grid with carbon film, using the process described in Ref [40]. The pentacene crystalline films were grown for 3 hours under the source temperature of 160 °C, which ~5-10 layers were typically produced. As the pentacene crystals would be damaged in tens of seconds when exposed under several hundreds of electron counts per squared nanometers, SAED patterns were acquired with an exposure time of 0.5s at a dose rate of 300 electrons per squared nanometers per second. However, the exposure time was not enough to collect HRTEM of pentacene crystals before the samples were damaged.

Raman spectroscopy was performed by a WITec Alpha 300R confocal Raman microscope with a 532nm laser (spot size~300nm, laser power 200pW).

Polarization-dependent absorption measurement was performed on the same WITec system with two linear polarizers (one between illumination source and sample, and the other between sample and detector), and without the notch filter. The two polarizers were approximately cross-polarized to minimize the background signal of SiO$_2$, since the absorption by 1L or 2L was very weak. White light was illuminated on the sample through a 50× objective lens, and the reflected light was collected by a CCD camera through a spectrometer. The images were obtained by scanning the sample with 300nm per step and 0.5s integration time. We plotted the images by integrating the spectrum from 520 to 550nm. Since the signal from the pentacene/BN was always lower than that from SiO$_2$/Si substrate, the collected image represented the absorption of the pentacene crystals.

For the polarization-dependent PL experiment, the sample was excited by a 488nm laser and a half-wave plate was used to allow a controllable change of the polarization of the laser. The fluorescence signal was collected by a 50× microscope objective and sent through a 0.5m spectrometer to a CCD camera.

**Fabrication process of OFETs and electrical measurements**

After growth, careful AFM screening was done prior to device fabrication to identify highly smooth and uniform samples with desirable number of layers for transport measurements. Two Au films were carefully transferred on the top of the crystals as source and drain electrodes under a microscope. The highly conductive silicon was used as global backgate. Electrical measurements were performed by a Keithley 4200 semiconductor parameter analyzer in a close-cycle cryogenic probe station under vacuum (~10$^{-5}$ Torr).

The transfer process of Au electrodes is described in detail here (Fig. S1). We first deposited 100nm Au film on SiO$_2$/Si substrates and scratch the film into rectangular patches with ~150μm × 50μm in size, using a tungsten tip attached to a micromanipulator. The transfer of much larger patches resulted in lower yield. We dipped the tungsten tip into gallium-indium eutectic as glue, and carefully picked up a patch from the corner under a microscope. We then used the micromanipulator to bring the Au patch on the top of the target area with pentacene on BN, and carefully lowered the probe tip, such that one side of the Au patch landed first, followed by the other side. The process was repeated twice for source and drain electrodes. No post-annealing was needed. This method gives conformal contact between the ultra-flat pentacene and Au, as reflected from the Ohmic contact of the devices. We are able to align the electrodes with micron precision, as reflected from the source-drain distance of our devices. The whole transfer process is carried out in ambient. During the transfer, it is important to minimize the vibration of the surroundings to achieve the best alignment and contact quality.

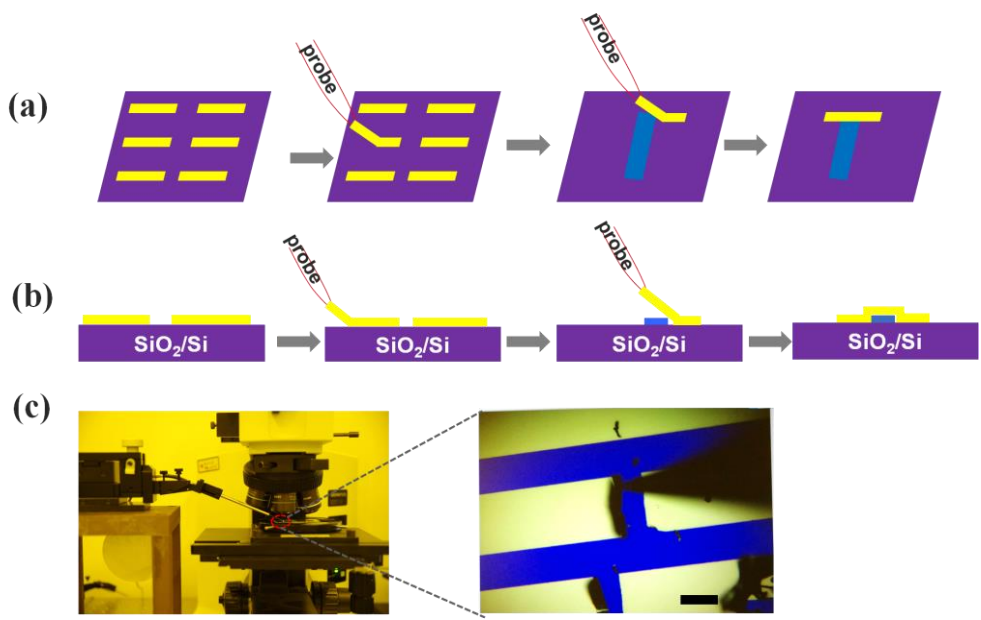

**Figure S1.** Cartoon illustration of the transfer process of Au electrodes, top view (a) and side view (b). (c) Picture of the transfer setup (left) and the microscope image of the tungsten tip picking up the Au film on SiO$_2$/Si substrate (right). The scale bar is 30μm.

**Fitting procedure of the hopping transport**

In the Vissenberg-Matters model [33], the conductivity is a function of the carrier density and temperature: $\sigma(p) = A p^{T_0/T}$, where $p$ is the carrier density in OFETs, $T_0$ is the width of the exponential distribution of the localized tail states, and

$$A = \sigma_0 \left( \frac{\left(\frac{T_0}{T}\right)^4 \sin\left(\pi \frac{T}{T_0}\right)}{(2\alpha)^3 B_c} \right)^{T_0/T} \quad (1)$$

In this equation, $\sigma_0$ is the conductivity prefactor, $\alpha^{-1}$ is the localization length of the wave function, and $B_c$ is the critical number for the onset of percolation ($B_c$=2.8 and 4.48 for 3D and 2D systems respectively) [41]. In 2D systems, the source-drain current can be presented as [34]

$$I_{ds}^{2D} = A \frac{W}{L} (d_{sc})^{1-(T_0/T)} \left(\frac{C_i}{e}\right)^{T_0/T} \frac{T}{T_0+T} \left[ (V_{th} - V_g)^{(T_0/T)+1} - (V_{th} - V_g + V_{ds})^{(T_0/T)+1} \right] \quad (2),$$

where $d_{sc}$ is the semiconductor thickness, $W$ and $L$ are the width and length of the transistor channel, respectively, $C_i$ is the gate capacitance per unit area, $V_{th}$ is the threshold voltage. Using Taylor expansion, the current at high gate bias reduces to a simple power law,

$$I_{ds}^{2D} \approx (V_{th} - V_g)^{T_0/T}. \quad (3)$$

The power exponents at different temperatures can be extracted from $I_{ds}$-$V_g$ characteristics plotted on a double logarithmic scale (Fig. 3b). By linearly fitting the function of power exponent as a function of $1/T$, we can prove that 2D hopping model is applicable in our devices and obtain $T_0$ from the slope. By applying the obtained $T_0$ and $d_{sc}$=1.2nm (obtained from AFM measurements) to Eq. 2, the $I_{ds}$-$V_g$ curves at different temperatures can be fitted by utilizing a single set of parameters of $\sigma_0$ and $\alpha^{-1}$. Using the same parameters, the field-effect mobility is then calculated as (Fig. 3c):

$$\mu = \frac{L}{C_i W V_{ds}} \frac{\partial I_{ds}}{\partial V_g} = \frac{A}{C_i V_{ds}} (d_{sc})^{1-(T_0/T)} (\frac{C_i}{e})^{T_0/T} [(V_{th} - V_g + V_{ds})^{T_0/T} - (V_{th} - V_g)^{(T_0/T)}]. \quad (4)$$

We note that the hopping picture holds when $T$ is smaller than $T_0$. In the high temperature limit (i.e. when $T \approx T_0$), all the molecular states are thermally accessible, linear $I_{ds}$-$V_g$ characteristics and carrier-density-independent mobility are recovered (Fig. 3a).

**General methods of DFT calculations**

DFT calculations were performed using the generalized gradient approximation for the exchange-correlation potential, the projector augmented wave method [42, 43], and a plane-wave basis set as implemented in the Vienna *ab-initio* simulation package. [44] The energy cutoff for the plane-wave basis was set to 400 eV for all calculations. In structural optimization, van der Waals interactions were considered in the framework of vdW-DF [45, 46] with the optB86b functional for the exchange energy (optB86b-vdW) [47,48]. Electronic band structure and wave function visualization were calculated with the same functional of structural relaxation, namely optB86b-vdW, and double checked with the Heyd-Scuseria-Ernzerhof screened hybrid functional (HSE06) [49,50]. Charge transfer integrals and reorganization energy were revealed by the HSE06 method based on the atomic structures fully optimized with the optB86b-vdW functional.

**Modeling the structure of epitaxial pentacene on BN**

To find the adsorption site and orientation of the WL pentacene molecules on BN, we put one molecule onto a bilayer BN in a 10×6 supercell. Only the Gamma point was adopted to sample the first surface Brillouin zone (BZ). During the structural relaxation, all atoms were fully relaxed until the residual force per atom was less than 0.01 eV·Å$^{-1}$, except the bottom layer of BN. We only focus on the lying-down configuration for WL in this work based on experimental observations. We examined in total 16 different configurations with four adsorption sites and four orientations.

The structure of 1L and 2L is incommensurate with WL. To simplify the model for the multiple-layer structure, a $\begin{pmatrix} 3 & 1 \\ 1 & 6 \end{pmatrix}$ BN supercell, with $a$ = 6.04 Å, $b$ = 15.55 Å and $\gamma$ = 90° was used, which contained one molecule in WL, four molecules in 1L and four molecules in 2L (Supplementary Fig. S17). A $k$-mesh of 4×2×1 was adopted to sample the surface BZ. The relative position of WL and 1L were determined from totally 24 relative configurations, including the consideration of different orientations, relative sites, tilting angles along $a$ and $b$. A series of lattice constant sets within the range of 5.93 ~ 6.50 Å for $a$-axis and 7.29 ~ 8.01 Å for $b$-axis were used to find the optimized lattice parameter for 1L and 2L (Supplementary Fig. S16). In each case, the shape and volume of the supercell were fixed and all atoms in the supercell were allowed to fully relax until the residual force per atom was less than 0.01 eV·Å$^{-1}$, except the WL molecule adopted directly from the WL structure. Since the BN sheet appeared to have nearly no effect on the structure of 1L and 2L, it was not included in the lattice optimization. Minimum external stress along both *a*- and *b*-axis was found with $a \approx 6.20$ Å and $b \approx 7.65$ Å for 1L (Supplementary Fig. S16), in excellent agreement with experimental values (less than 1.5% mismatch). We found that the electronic structures did not change significantly upon small structural changes (e.g., 1.5%) for both 1L and 2L. We thus used the experimental lattice parameters from AFM measurements in our further analysis of the electronic structures of the 1L and 2L.

To avoid very large supercells in DFT calculations, we have to slightly compress the BN sheet by 6.3% along *x*-axis and expand the *y*-axis by 10% to accommodate the four molecules for 1L. This compression leads to relatively larger electronic band dispersion within WL, but keeps the electronic structures of 1L and 2L nearly unaffected. We have carefully checked this model with a larger supercell, namely three molecules in WL, 14 molecules in 1L and another 14 molecules in 2L. This larger supercell essentially gives the same results to the simpler model and reassures the validity of our calculations.

Inter- and intra-layer interactions for the multilayer thin films were calculated with the following formulas.

$$E_{\text{BN}-\text{WL}}^{\text{inter}} = E_{\text{BN}+\text{WL}} - E_{\text{BN}} - E_{\text{Mol}} \quad (5)$$

$$E_{\text{WL}-1\text{L}}^{\text{inter}} = \frac{1}{4}(E_{\text{Total}} - E_{\text{WL}} - E_{1\text{L}+2\text{L}}) \quad (6)$$

$$E_{1\text{L}-2\text{L}}^{\text{inter}} = \frac{1}{4}(E_{\text{Total}} - E_{\text{WL}+1\text{L}} - E_{2\text{L}}) \quad (7)$$

$$E_{\text{WL}}^{\text{intra}} = \frac{1}{2}(E_{WL} - E_{\text{Mol}} \times 2) \qquad (8)$$

$$E_{\text{1L}}^{\text{intra}} = \frac{1}{4}(E_{1L} - E_{\text{Mol}} \times 4) \qquad (9)$$

$$E_{\text{2L}}^{\text{intra}} = \frac{1}{4}(E_{2L} - E_{\text{Mol}} \times 4) \qquad (10)$$

Here, $E_{\text{Total}}$, $E_{\text{WL}}$, $E_{\text{1L}}$, $E_{\text{2L}}$, $E_{\text{WL+1L}}$ and $E_{\text{1L+2L}}$ are the total energies of the WL+1L+2L, WL, 1L, 2L, WL+1L and 1L+2L thin films, and $E_{\text{Mol}}$ is the total energy of a single pentacene molecule in the gas phase. Derived interlayer interaction energies $E_{\text{BN-WL}}^{\text{inter}}$, $E_{\text{WL-1L}}^{\text{inter}}$ and $E_{\text{1L-2L}}^{\text{inter}}$ represent the BN-WL, WL-1L and 1L-2L interactions on the per molecule basis, respectively. Intra-layer interaction energies $E_{\text{WL}}^{\text{intra}}$, $E_{\text{1L}}^{\text{intra}}$ and $E_{\text{2L}}^{\text{intra}}$ indicate the molecule-molecule interaction within WL, 1L and 2L, respectively.

**Calculation of charge transfer integral**

Charge transfer integral is a key physical quantity to describe the charge transport in organic semiconductors [51],

$$V_{ij} = t_{ij} = \frac{1}{2}(E_{\text{HOMO}} - E_{\text{HOMO}-1}), \qquad (11)$$

where $E_{\text{HOMO}}$ and $E_{\text{HOMO}-1}$ are the two highest occupied molecular orbitals of a pentacene dimer. A large supercell, with at least 10 Å vacuum region in all the three directions, was adopted for each calculation of charge transfer integral. The atomic structures of molecular dimers for calculating charge transfer integral were directly taken from the optimized structure in WL, 1L and 2L. Energetic differences for extracting the charge transfer integral were computed using the HSE06 functional.

**2. Growth process of few-layer pentacene crystals on BN**

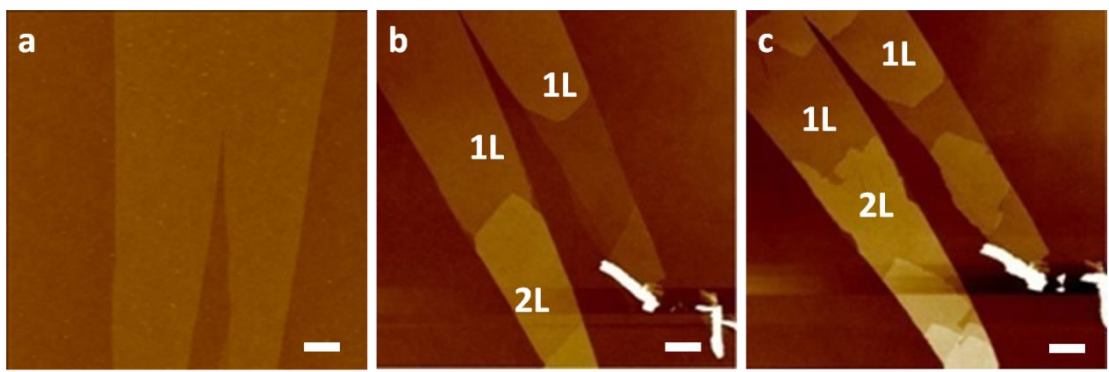

**Figure S2. Sequential AFM images of the growth process of pentacene on BN. a**, AFM of the BN sheet before growth. **b-c**, are the AFM images after 120mins and 150mins growth, respectively. The scale bars are 1μm. The layer numbers are marked on each image. The growth was conducted under high vacuum ~$4\times10^{-6}$ Torr without carrier gas and the temperature of the pentacene source was 130 °C.

3. **High-resolution AFM characterization of 2L pentacene crystal**

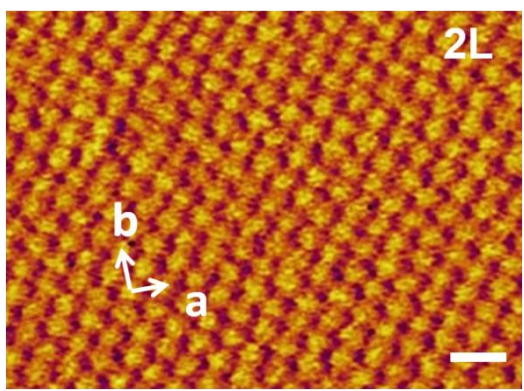

**Figure S3.** High-resolution AFM image of a 2L pentacene sample. The unit cell is marked. The scale bar is 1nm.

4. **TEM characterization of few-layer pentacene crystal**

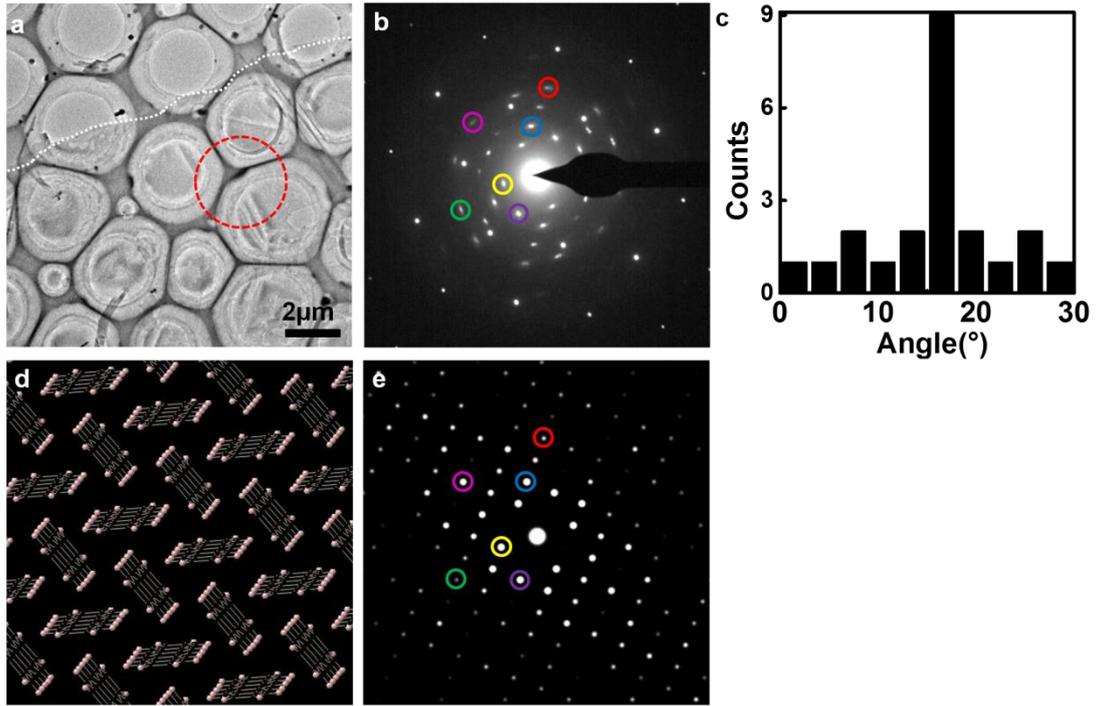

**Figure S4. a**, Low-magnification TEM image of a few-layer pentacene/BN sample. The white dotted line indicates the edge of the sample. The scale bar is 2μm. **b**, SAED pattern taken from the area marked by the red circle in **a**. **c**, Histogram distribution of the intersection angle between the crystal orientation of pentacene (010) and BN (100), taken from several different samples. The distribution shows a sharp peak near 16°, which proves that the pentacene has an epitaxial relationship with the BN sheet. However, the distributions in other angles suggest that the epitaxy is not fully fulfilled, due to its weak vdW nature. **d**, Structural model of pentacene crystal. The structure parameters come from Ref. [52]. **e**, Simulated diffraction patterns using the structure in **d**. The circles with the same color in **b** and **e** correspond to the same diffraction spots, as following: yellow (110), violet (020), green (230), purple ($3\bar{1}0$), blue ($1\bar{2}0$), red ($1\bar{4}0$). By utilizing the diffraction spots of BN as references, the lattice constants of pentacene

crystal can be calculated as 5.98Å±0.09Å and 7.61Å±0.13Å, respectively. The angle between *a*- and *b*-axis is 88.25 °±1.22 °. We note that the lattice constants from TEM and AFM differ by less than 2% due to system errors. However, this does not affect the main conclusion of our paper, which is the modulation of charge transport by molecular packing. The different molecular packing between 1L and 2L is unambiguously measured by AFM.

## 5. Polarization-dependent absorption and PL of few-layer pentacene crystal

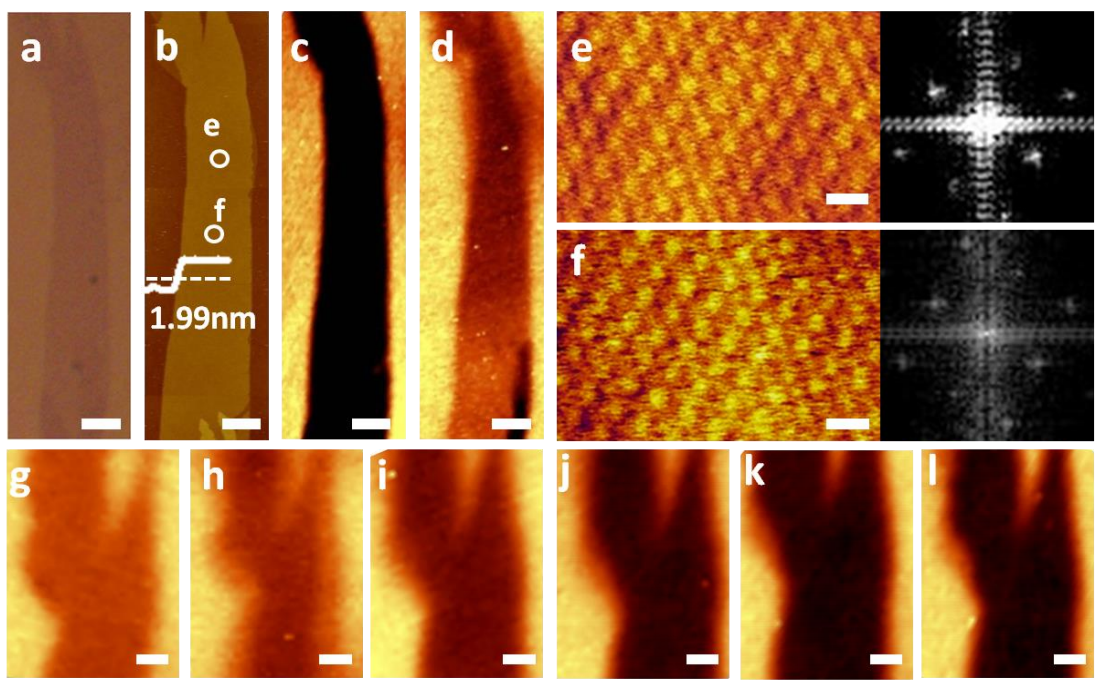

**Figure S5. Characterizations of a 1L pentacene sample on a monolayer BN with size of ~60μm. a**, Optical microscopy and **b**, AFM image of the sample after pentacene growth. The step height of 1.99nm suggests that the whole sample is covered by 1L pentacene crystals. **c** and **d** are polarization-dependent absorption microscopy images,

showing clear anisotropy as we rotate the sample by 45 °. The sample area is darker than the surrounding SiO$_2$/Si area. The scale bars in **a** to **d** are 5μm. The left panels of **e** and **f** are high-resolution AFM images taken from two randomly selected spots marked in **b**, clearly showing the same crystal orientation. Right panels are the Fast Fourier Transform of the AFM images. The scale bars in **e** and **f** are 1nm. From the AFM and polarization-dependent absorption microscopy images, we confirm that the entire sample is covered by single-crystalline 1L pentacene. From **g** to **l** are sequential images of polarization-dependent absorption microscopy from a portion of the sample in **a**. The scale bars are 2μm. With respect to **g**, the sample is rotated by 16 °(**h**), 32 °(**i**), 49 °(**j**), 62 °(**k**), and 78 °(**l**), respectively. As we rotate the sample, the absorption is uniformly modulated. The angle between the highest and lowest absorption is ~80 °, indicating that the anisotropy in absorption is tailored by the crystal symmetry of pentacene.

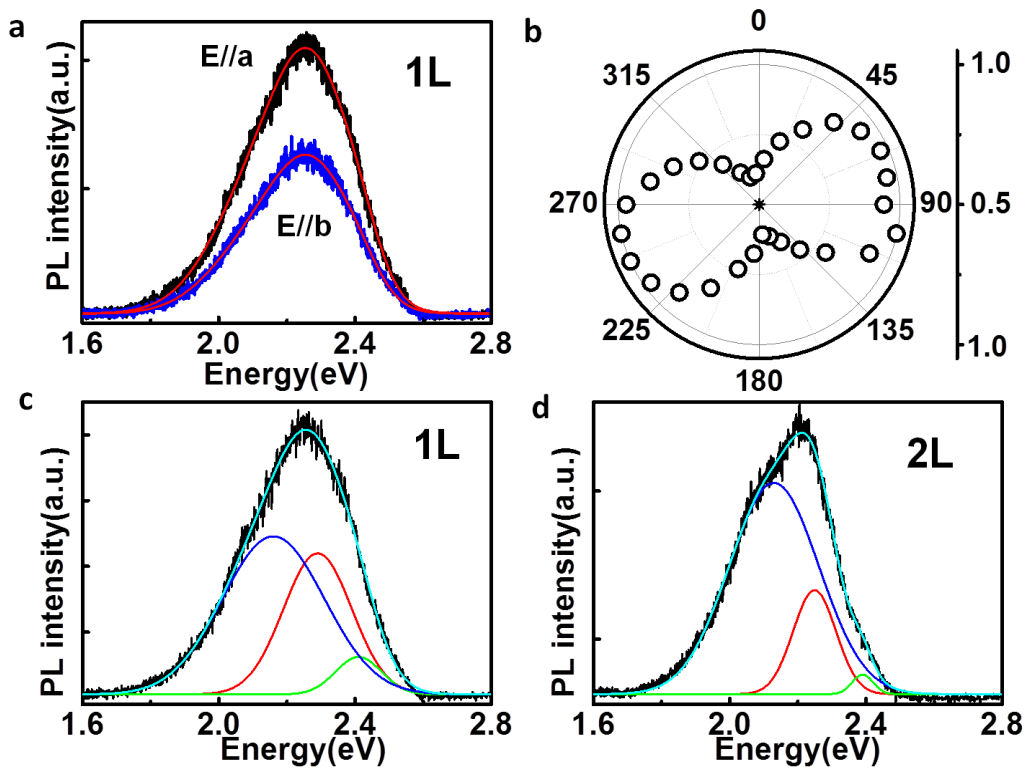

**Figure S6. Polarization-dependent PL measurements. a**, Polarization-dependent PL spectra from a 1L sample. The black and blue curves represent the PL spectra when the polarization direction is along *a*- and *b*-axes, respectively. Red lines are the fitting results with three Gaussian peaks. **b**, Normalized PL intensity of the 1L sample in **a** as a function of linear polarization angle. **c**, PL spectrum (black) and fitting (cyan) of the 1L pentacene crystal. Blue, red, and green lines are the three de-convoluted Gaussian peaks. The peak positions are at 2.16eV, 2.29eV and 2.41eV, respectively. **d**, PL spectrum (black) and fitting (cyan) of 2L pentacene crystals. Blue, red, and green lines are the three de-convoluted Gaussian peaks. The peak positions are at 2.13eV, 2.25eV and 2.39eV, respectively.

6. **Electrical data of few-layer pentacene FETs**

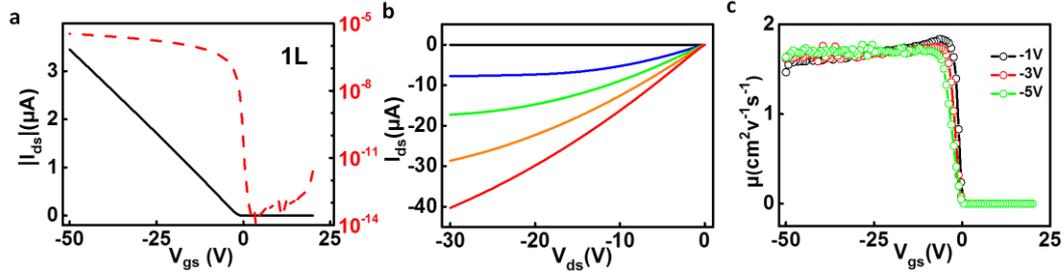

**Figure S7. Additional electrical data of the 1L pentacene device shown in Fig. 3a of the main text. a**, Room-temperature $I_{ds}$-$V_g$ characteristics ($V_{ds}=-2$V) in linear (black line) and log (red dash line) scale. For this device, $V_{th}=-1.2$V, and the subthreshold swing $SS=450$mV/decade. **b**, Room-temperature $I_{ds}$-$V_{ds}$ characteristics of the device. From top to bottom, $V_g=0$V, $-20$V, $-30$V, $-40$V, and $-50$V, respectively. **c**, The room-temperature $\mu$-$V_g$ relationship (in the linear regime) under different bias voltages. The mobility values at different biases are the same, indicating the contact resistance is negligible compared to the channel resistance.

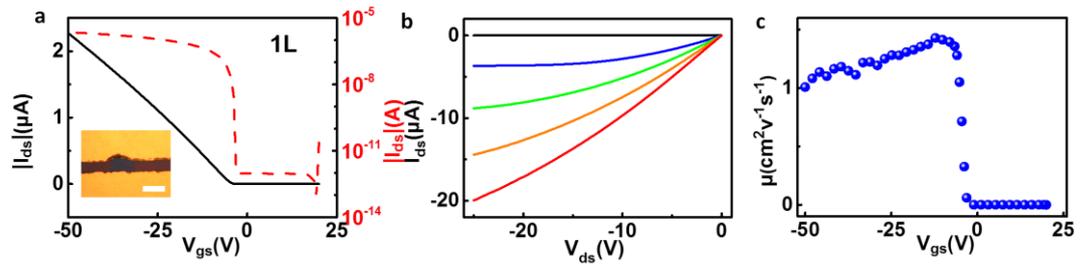

**Figure S8. Room-temperature electrical data of another 1L pentacene device. a**, $I_{ds}$-$V_g$ characteristics ($V_{ds}=-2$V) in linear (black line) and log (red dash) scale. Inset shows the optical microscopy image of the device. The scale bar is 10μm. For this device, $V_{th}=-2.84$V, and $SS=200$mV/decade. **b**, $I_{ds}$-$V_{ds}$ characteristics of the device.

From top to bottom, $V_g$=0V, −20V, −30V, −40V and −50V, respectively. **c**, The extracted $\mu$-$V_g$ relationship. The peak mobility is ~1.5 cm$^2$/Vs.

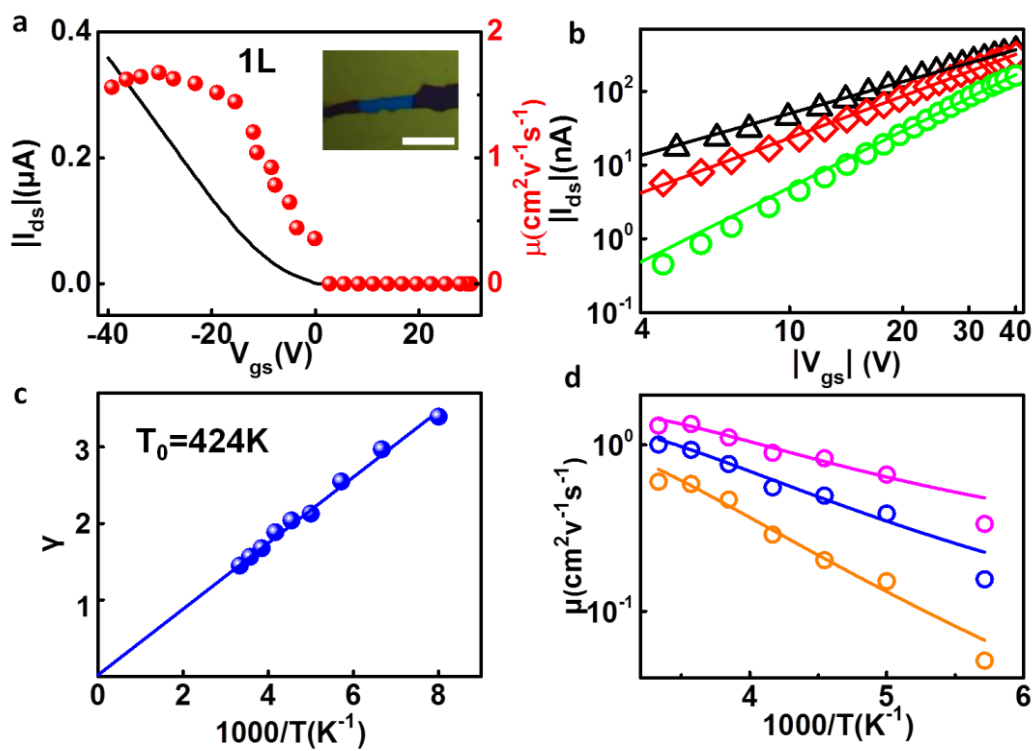

**Figure S9. Temperature-dependent electrical data of another 1L pentacene device.**
**a**, Room-temperature $I_{ds}$-$V_g$ characteristics ($V_{ds}$=−0.3V, black line), and the extracted $\mu$-$V_g$ relationship (red symbols). Inset shows the optical microscope image of the device. The scale bar is 10μm. The peak mobility is over 1.5 cm$^2$/Vs for this device. **b**, Experimental $I_{ds}$-$V_g$ characteristics under $V_{ds}$=0.3V (symbols) plotted on a double logarithmic scale at different temperatures. From top to bottom, $T$=300K, 240K and 175K, respectively. The lines are power-law fittings, showing excellent agreement over extended conductance and temperature range. **c**, The power exponent extracted from **b** as a function of 1000/$T$ (symbols). The linear fitting (line) crosses the origin. From the slope of the line, we obtain $T_0$=424K. **d**, Experimental (symbols) and calculated (lines) mobility as a function of 1000/$T$ at different gate voltages. From top to bottom, $V_g$=−15V, −10V and −5V, respectively. The calculations are done using the procedures described in the Methods section with the following parameters: $T_0$=424K, $\sigma_0$=4×10$^6$S/m, $\alpha^{-1}$=9.4Å.

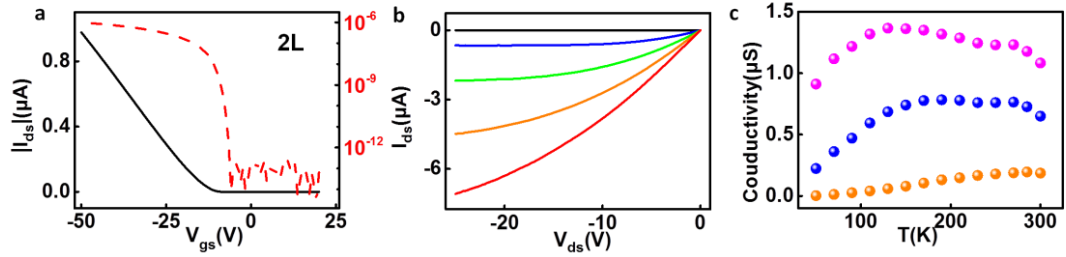

**Figure S10. Additional electrical data of the 2L pentacene device shown in Fig. 3d of the main text. a**, Room-temperature $I_{ds}$-$V_g$ characteristics ($V_{ds}$=−2V) in linear (black) and log (red dash) scale. **b**, Room-temperature $I_{ds}$-$V_{ds}$ characteristics of the device. From top to bottom, $V_g$=−10V, −20V, −30V, −40V and −50V, respectively. **c**, Conductivity as a function of temperature under $V_g$=−20V (orange), −35V (blue) and −50V (purple).

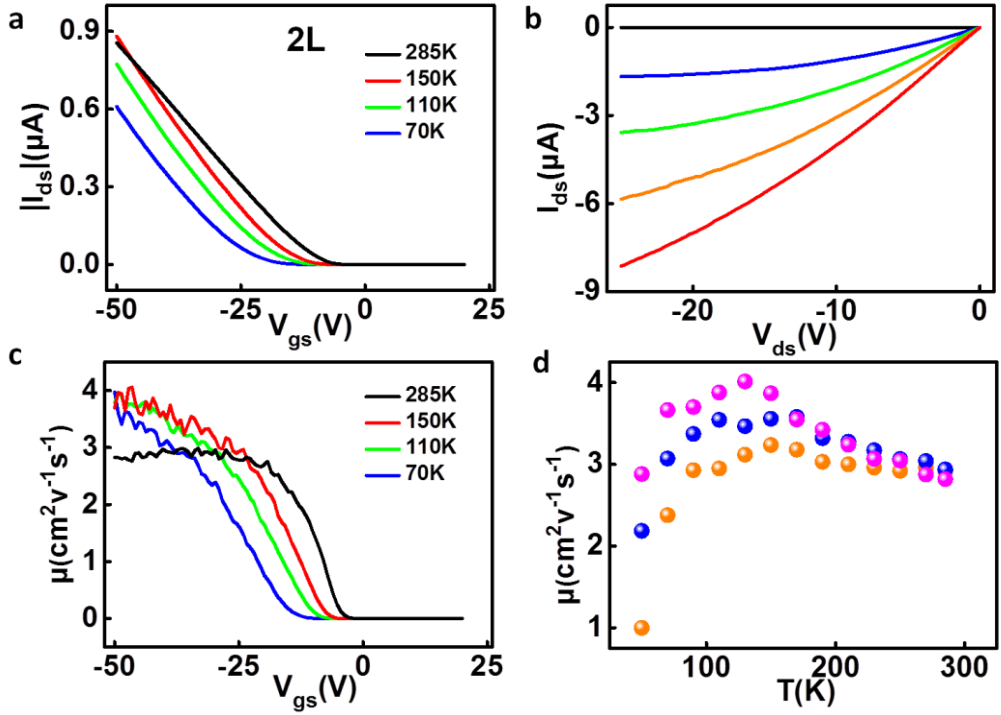

**Figure S11. Temperature-dependent electrical data of another 2L pentacene device. a**, $I_{ds}$-$V_g$ characteristics under $V_{ds}$=−2V at different temperatures. Black, red, green and blue lines represent $T$=285K, 150K, 110K and 70K, respectively. **b**, Room-temperature $I_{ds}$-$V_{ds}$ characteristics of the device. From top to bottom, $V_g$=−10V, −20V, −30V, −40V, and −50V, respectively. **c**, The extracted $\mu$-$V_g$ relationship at the same temperatures as in **a**. The room-temperature mobility reaches ~3cm$^2$/Vs in this device. **d**, Mobility as a function of temperature under $V_g$=−30V (orange), −40V (blue), and −50V (purple). Metallic behavior is observed down to ~150K.

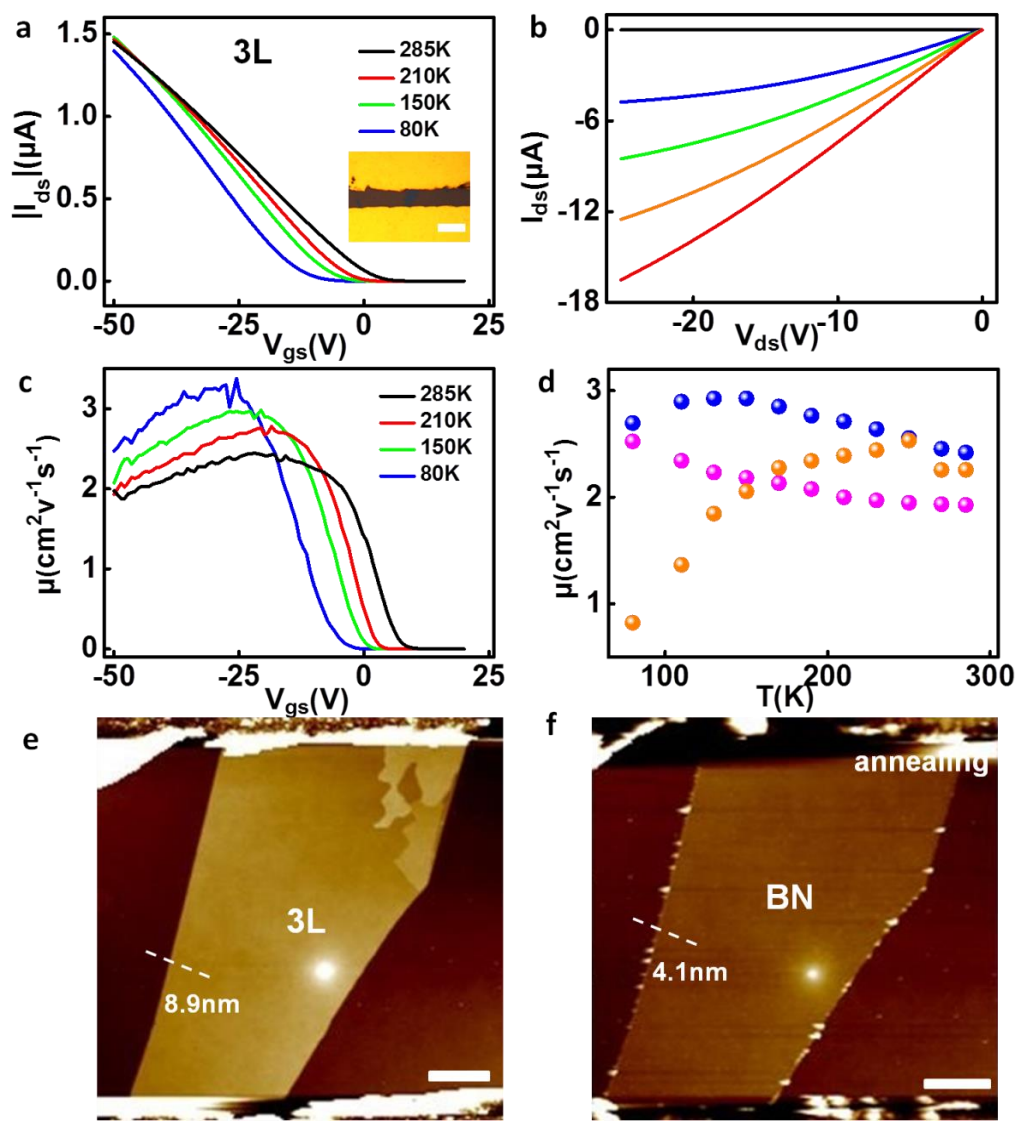

**Figure S12. Temperature-dependent electrical data of a 3L pentacene device. a**, $I_{ds}$-$V_g$ characteristics under $V_{ds}$=−2V at different temperatures. Black, red, green and blue lines represent $T$=285K, 210K, 150K and 80K, respectively. Inset shows the optical microscope image of the device. The scale bar is 10μm. **b**, Room-temperature $I_{ds}$-$V_{ds}$ characteristics of the device. From top to bottom, $V_g$=−10V, −20V, −30V, −40V and −50V, respectively. **c**, The extracted $\mu$-$V_g$ relationship at the same temperatures as in **a**. **d**, Mobility as a function of temperature under $V_g$=−10V (orange), −20V (blue),

and −50V (purple). Metallic behavior is observed down to sub-100K. **e**, AFM image of the 3L device. The total height of the sample is 8.9nm (including the height of BN and pentacene crystals). We note that complete 3L coverage is not easy to control due to sample-to-sample variations. In many cases, a small portion of 4L appears after the completion of 3L, as in the case of this device. However, we do not expect the 4L to significantly affect the charge transport due to its small size. In order to confirm that the pentacene has 3L, after all the measurements, we heated the sample to 500℃ for an hour in high vacuum to remove the pentacene, and measure the height of the BN sheet. **f**, shows the AFM image of the device after the pentacene removed. The height measurement indicates that the total thickness of the pentacene is 4.8nm, which is consistent with 3L sample. The scale bar in **e** and **f** is 10μm.

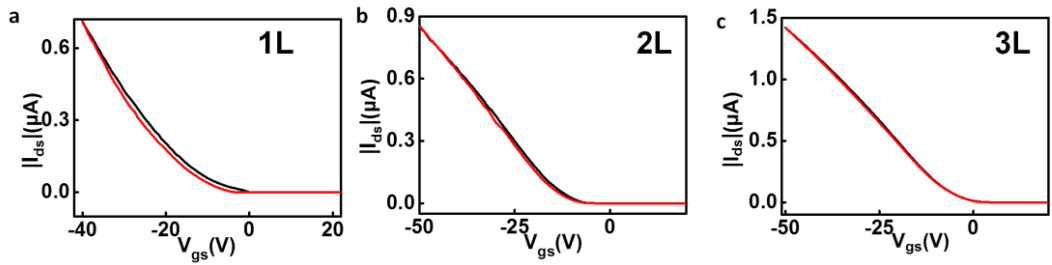

**Figure S13. Hysteresis of pentacene FETs. a**, Room-temperature double sweep $I_{ds}$-$V_g$ characteristics ($V_{ds}$=−0.7V) of the 1L device shown in Supplementary Fig. 9. **b**, Room-temperature double sweep $I_{ds}$-$V_g$ characteristics ($V_{ds}$=−2V) of the 2L device shown in Fig. 3d of the main text. **c**, Room-temperature double sweep $I_{ds}$-$V_g$ characteristics ($V_{ds}$=−2V) of the 3L device shown in Supplementary Fig.12. All devices show the negligible hysteresis.

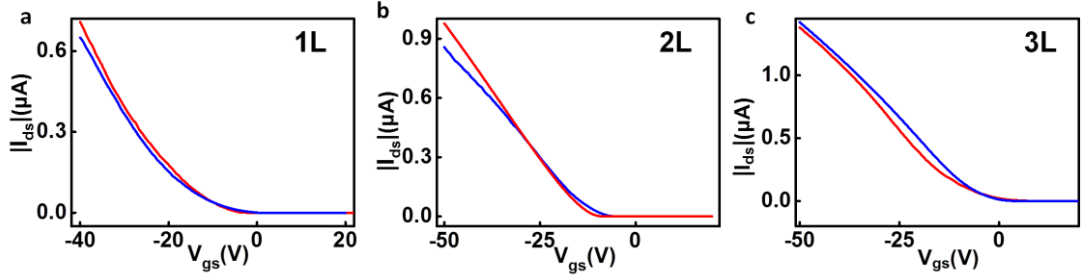

**Figure S14. Repeatability of pentacene FETs.** Room temperature $I_{ds}$-$V_g$ characteristics of the 1L device in Supplementary Fig. 9 (**a**), the 2L device in Fig. 3 of the main text (**b**), and the 3L device in Supplementary Fig.12 (**c**). Red and blue lines are measured before cooling down and after the complete thermal cycle to the base temperature, respectively. The reversible electrical characteristics suggest that the samples can be fully recovered after a thermal cycle, and that our observations are not due to any irreversible damage of the samples.

7. **Geometric structure of epitaxial pentacene determined by DFT calculations.**

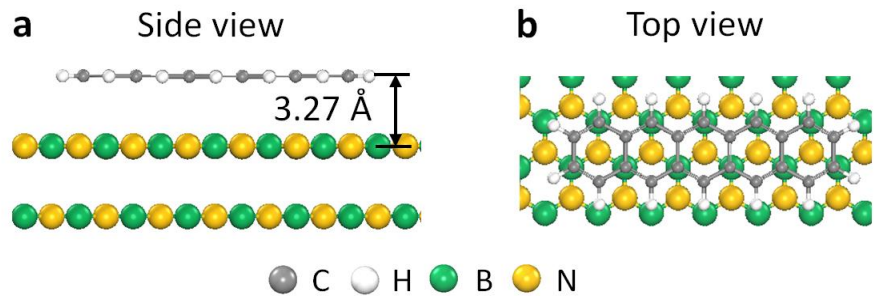

**Figure S15. a**, Side view and **b**, top view of the most energetically favorable configuration of WL pentacene molecules on BN. The pentacene molecule adopts the face-on configuration so that the phenyl rings of the molecule are orientated parallel to the BN sheet. The adsorption distance is 3.27 Å and the molecular longer axis is along the [11$\bar{2}$] direction of the BN sheet with a relative interlayer position similar to that of Bernal-stacking. This configuration shares the same structural features of previously reported adsorption configuration of pentacene adsorbed on graphene [5].

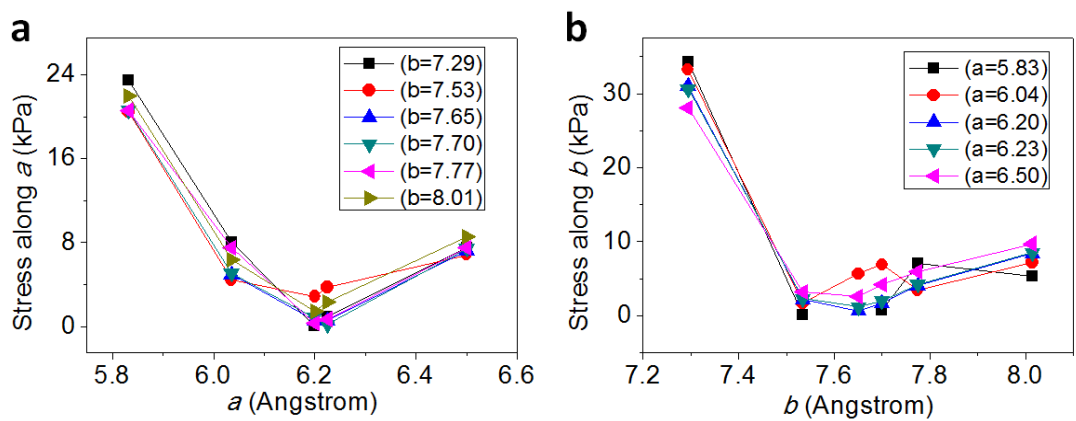

**Figure S16.** Optimization of lattice parameters of 1L pentacene. The lattice stress of the supercell with different combinations of *a* and *b* is presented here. **a**, External stress along *a*-axis as a function of *a*, while *b* is fixed. The corresponding *b* for each curve is shown. **b**, External stress along *b*-axis as a function of *b*, while *a* is fixed. The corresponding *a* for each curve is shown. From both panels, we find the overall minimum stress is obtained with $a \approx 6.20$ Å and $b \approx 7.65$ Å. We note that there is a slight mismatch under 1.5% between the lattice constants measured by AFM and optimized by DFT calculations. However, the electronic structure and wave function distribution do not change significantly upon such small structural changes. We therefore use the lattice parameters from AFM measurements to perform the DFT calculations and focus our discussions on the difference between 1L and 2L induced by the structural change.

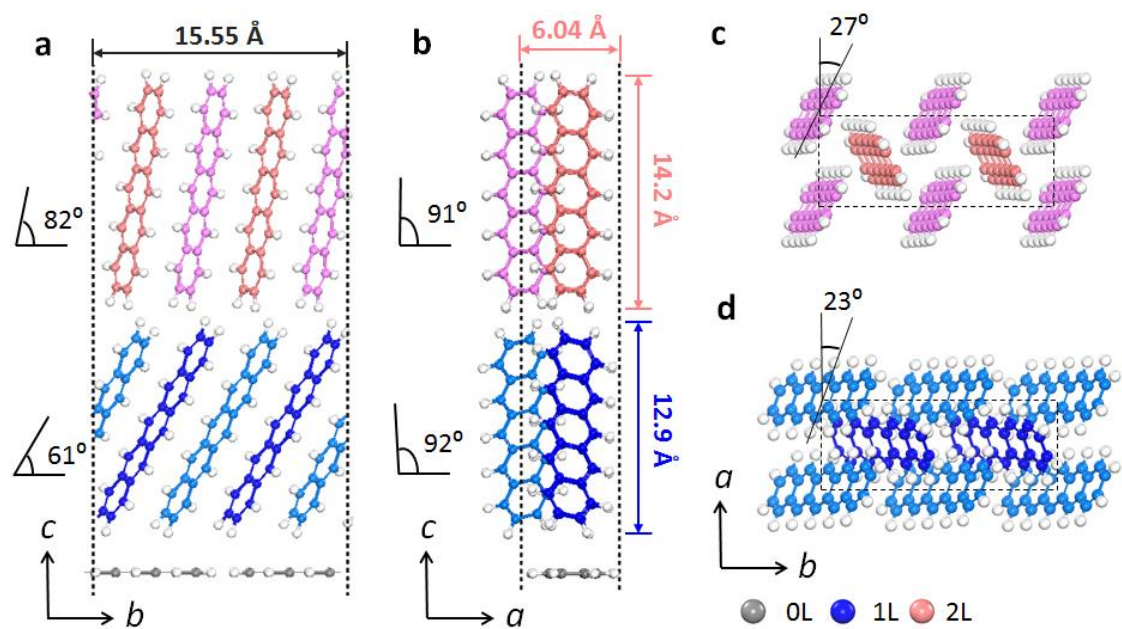

**Figure S17. Optimized geometric structure of the WL+1L+2L pentacene crystal.** Side views of the crystal along the *b*- and *a*-axes are shown in **a** and **b**, respectively. **c**, and **d** show the molecular structure of 2L and 1L projected into the *a-b* plane. Lattice parameters *a* and *b*, the tilting angles and the vertical distances of the molecular skeleton are marked. The tilting angles are 61 ° and 82 ° for the 1L and 2L molecules with respect to the *b*-axis; while they are close to 90 ° with respect to the *a*-axis. The reduced tilting angle of 2L makes the molecules reorient their shorter axis more off the *a*-axis, namely from 23 ° in 1L to 27 ° in 2L. The increased angle results in larger intermolecular spacing (thus more attraction) along the *a*-axis but smaller (thus more repulsion) for the *b*-axis. The repulsion along the *b*-axis is largely cancelled by the attraction led by less tilted 2L molecules. Therefore, the lattice of 2L shrinks along *a*-axis, but is nearly unchanged along *b*-axis compared with that of 1L.

## 8. Electronic properties of epitaxial pentacene.

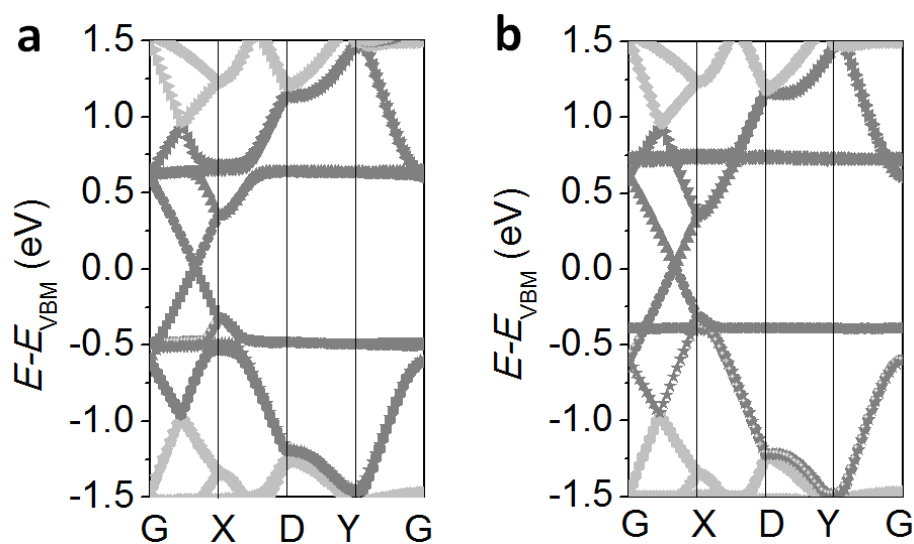

**Figure S18. Electronic band structures of pentacene adsorbed on *graphene*.** As a crosscheck of our calculation methods, we try to reproduce widely accepted results from the literature. **a**, The reported band structure of pentacene on graphene [53] and **b**, our results using the same structure as Ref.[53] . The main features of the band structures are successfully reproduced. We note that the structure of the interface was not relaxed in Ref. [53]. We find that the structural relaxation using an accurate optB86b-vdW functional as adopted in this work may lead to the molecules slightly tilted along their shorter axis. However, this change in geometry does not result in qualitative difference in the band structures as shown in **b**.

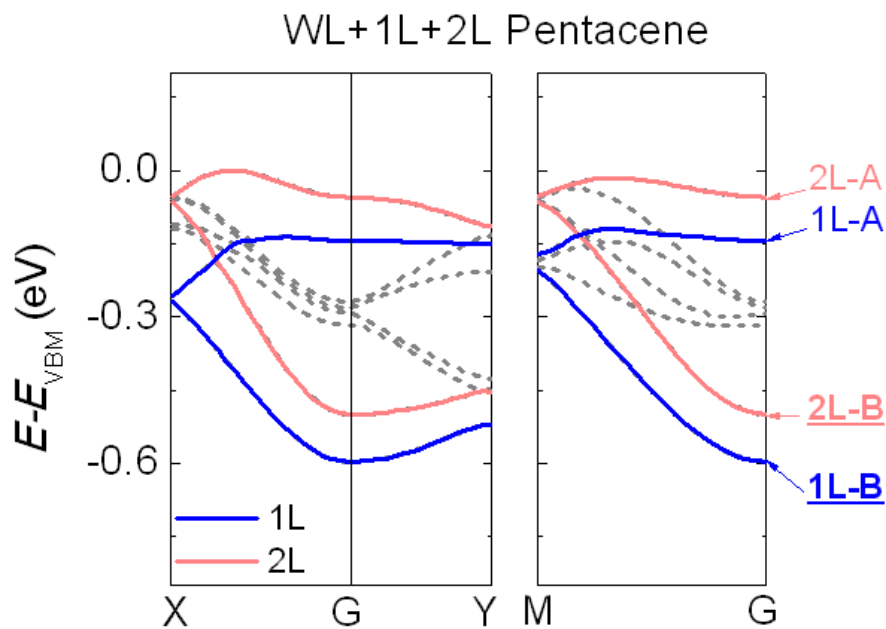

**Figure S19.** Band structures of the hybrid WL-1L-2L structure as shown in Supplementary Fig. 17. Blue and red solid lines represent the electronic states from 1L and 2L, respectively. We find four states, namely 2L-A (anti-bonding), 1L-A (anti-bonding), 2L-B (bonding) and 1L-B (bonding) are of particular interest. There are two non-equivalent pentacene molecules in a unit cell of 1L or 2L. Stacking them together into a layer results in a separation of the original HOMO, equivalently to form inter-molecular bonding (1L-B or 2L-B) and anti-bonding (1L-A or 2L-A) states. Here "anti-bonding" and "bonding" refer to the inter-molecular interactions. The assessment made here is also consistent with the fact that bonding states have lower eigen-energies than those of anti-bonding states (also see Supplementary Fig. 21).

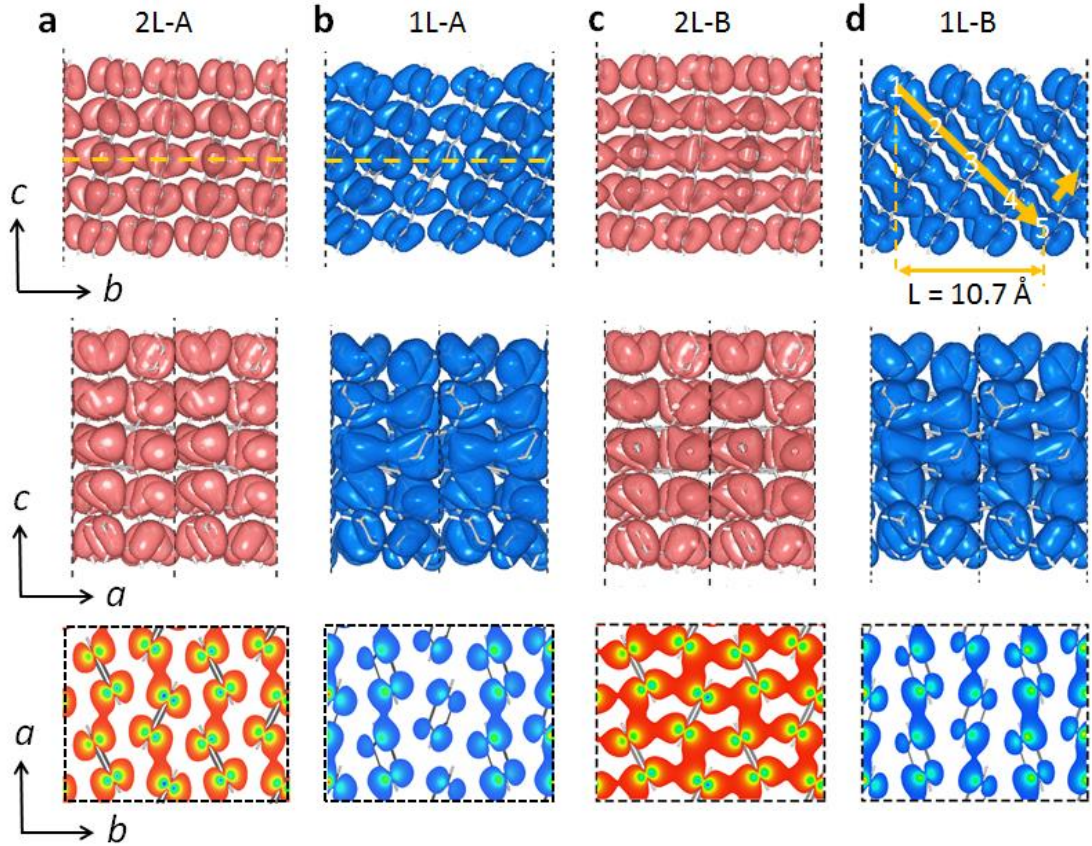

**Figure S20.** Visualized wave functions of the four states 2L-A (**a**), 1L-A (**b**), 2L-B (**c**) and 1L-B (**d**), illustrated by isosurface contours (0.00015 $e$/Bohr$^3$) in the *b-c* (top panel) and *a-c* planes (middle panel) and in the *a-b* plane of a slice (bottom panel). The position of the slice is indicated by the yellow dashed line in (**a**) and (**b**). We find that states 2L-B and 1L-B are extended in the real space and clearly have intermolecular wave function overlaps that 2L-A and 1L-A cannot offer. We therefore focus on states 1L-B and 2L-B in the main text, which, we believe, govern the current flow in the pentacene devices. Interestingly, the state 1L-B is partially extended so that the wave function only spans five molecules along the *b*-axis, which gives a localization length of roughly 1 nm.

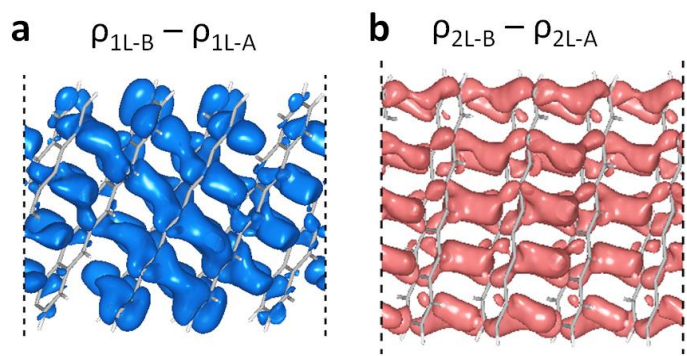

**Figure S21.** Differential charge density between 1L-B and 1L-A states (**a**) and 2L-B and 2L-A (**b**). The blue and red isosurface contours indicate the electron accumulation occurs spatially between the two adjacent molecules. These charge density mapping explicitly shows that 1L-B and 2L-B are intermolecular bonding states and 1L-A and 2L-A are anti-bonding states.

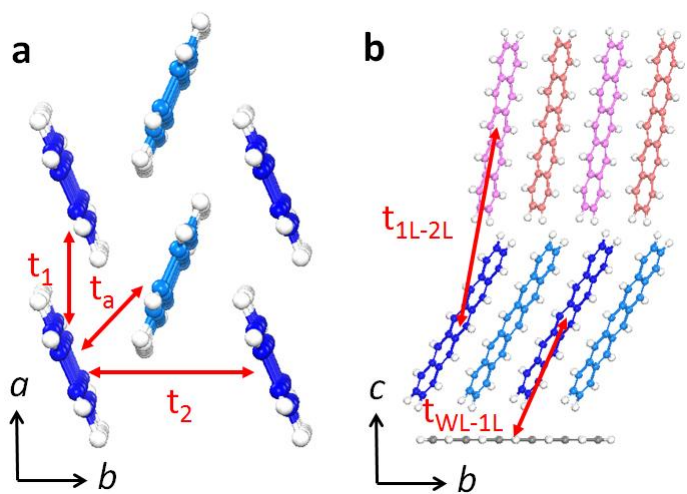

**Figure S22. Charge transfer integral is derived on the molecular dimer basis.** We consider three intra-layer dimer pairs in 1L shown in (**a**) and an inter-layer dimer pairs of 1L-2L and WL-1L, as shown in (**b**). All dimer pairs are marked with red arrows. Dimer pairs $t_1$, $t_2$ and $t_a$ represent charge transfer integrals along the [1,0] , [0,1]

and [1/2, ±1/2] directions, respectively. Interlayer coupling of $t_{\text{WL-1L}}$ and $t_{\text{1L-2L}}$ represent the charge transfer integrals between WL and 1L and between 1L and 2L.

| Structure | Inter-layer interaction | | | Intra-layer interaction | | |
|---|---|---|---|---|---|---|
| | $E_{\text{BN-WL}}^{\text{inter}}$ | $E_{\text{WL-1L}}^{\text{inter}}$ | $E_{\text{1L-2L}}^{\text{inter}}$ | $E_{\text{WL}}^{\text{intra}}$ | $E_{\text{1L}}^{\text{intra}}$ | $E_{\text{2L}}^{\text{intra}}$ |
| Energy (eV) | -2.35 | -0.26 | -0.28 | -0.32 | -1.72 | -1.79 |

**Table S1.** Inter- and intra-layer interaction energies for the multilayer thin films. For WL, the interlayer energy with the BN sheet $E_{\text{BN-WL}}^{\text{inter}}$ is the largest among all listed values, even higher than the sum of intra- and inter-layer interactions for 2L or 1L. Therefore, WL unambiguously adopts the lying-down configuration. The inter-layer interaction energy reduces to 0.2 to 0.3 eV for WL-1L and 1L-2L; while the intra-molecular interaction within a layer enhances to over 1.7 eV owing to strong π-π stacking in 1L and 2L.

| Transfer integrals (eV) | |
|---|---|
| $t_1$ [1,0] | 0.036 |
| $t_2$ [0,1] | 0.000 |
| $t_a$ [1/2,±1/2] | 0.175 |
| $t_{\text{WL-1L}}$ | 0.016 ~ 0.090 |
| $t_{\text{1L-2L}}$ | < 0.010 |

**Table S2.** Charge transfer integrals, $t_{ij}$, of the dimer pairs indicated in Supplementary Fig. 22. The charge transfer integrals are 0.036 eV, 0.000 eV and 0.175 eV for dimer pairs $t_1$, $t_2$ and $t_a$, which are inversely correlated with the inter-molecular distances. The largest charge transfer integral of 0.175 eV indicates that the [1/2, ±1/2] direction is the most likely direction for current flow. $t_{WL-1L}$ has a range of values, corresponding to different dimer pairs between WL and 1L, since WL is incommensurate with respect to 1L. $t_{WL-1L}$ is comparable to the intralayer counterparts in 1L, suggesting strong WL-1L electronic coupling. Therefore, carriers can easily hop from 1L to WL. Due to the lack of π-π stacking in WL, the carriers can be further localized. However, the coupling between 1L and 2L is negligible. Therefore, the presence of 1L does not affect the charge transport in 2L.